\documentclass[12pt]{iopart}

\usepackage{iopams}  
\usepackage{amstext,amssymb}
\usepackage[dvips]{graphicx}
\usepackage{cite}
\usepackage{xspace}

\newcommand{\dms}{\Delta m^2_{21}}
\newcommand{\dma}{\Delta m^2_{31}}
\newcommand{\eVq}  {\text{eV}^2}
\newcommand{\stheta}  {\sin^22\theta_{13}}

\newcommand{\nue}{\ensuremath{\nu_{e}}\xspace}
\newcommand{\numu}{\ensuremath{\nu_{\mu}}\xspace}
\newcommand{\nutau}{\ensuremath{\nu_{\tau}}\xspace}
\newcommand{\nubarmu}{\ensuremath{\overline{\nu}_{\mu}}\xspace}
\newcommand{\nubare}{\ensuremath{\overline{\nu}_{e}}\xspace}
\newcommand{\numunue}{\ensuremath{\nu_\mu \rightarrow \nu_e}\xspace}
\newcommand{\ceren}{\v{C}erenkov\xspace}
\newcommand{\thetaot}{\ensuremath{\theta_{13}}\xspace}
\newcommand{\delCP}{\ensuremath{\delta_{\rm CP}}\xspace}

\begin{document}

{\hfill EURONU-WP6-10-15}
\vspace*{1cm}

\topical{$\theta_{13}$: phenomenology, present status and prospect}

\author{Mauro Mezzetto}
\address{Instituto Nazionale Fisica Nucleare, Sezione di Padova,\\
  Via Marzolo 8, 35131 Padova, Italy}
\ead{mauro.mezzetto\_at\_pd.infn.it}

\author{Thomas Schwetz}
\address{Max-Planck-Institute for Nuclear Physics,\\
PO Box 103980, 69029 Heidelberg, Germany}
\ead{schwetz\_at\_mpi-hd.mpg.de}

\begin{abstract}
The leptonic mixing angle $\theta_{13}$ is currently a high-priority topic
in the field of neutrino physics, with five experiments under way,
searching for neutrino oscillations induced by this angle. We review the
phenomenology of $\theta_{13}$ and discuss the information from present
global oscillation data. A description of the upcoming reactor and
accelerator experiments searching for a non-zero value of $\theta_{13}$ is
given, and we evaluate the sensitivity reach within the next few years.
\end{abstract}


\maketitle

\tableofcontents

\title[$\theta_{13}$: phenomenology, present status and prospect]{}

\newpage
\section{Introduction}

Neutrino oscillations have been firmly established in the last twelve years
or so by a beautiful series of experiments with neutrinos from the
sun~\cite{Cleveland:1998nv, Altmann:2005ix, Hosaka:2005um, Ahmad:2002jz,
Aharmim:2008kc, Collaboration:2009gd, Arpesella:2008mt}, the Earth's
atmosphere~\cite{Fukuda:1998mi, Ashie:2005ik}, nuclear
reactors~\cite{Araki:2004mb, :2008ee}, and accelerators~\cite{Ahn:2006zz,
Adamson:2008zt}. All these data can be described within a three--flavour
neutrino oscillation framework, characterised by two mass-squared
differences ($\dms, \dma$), three mixing angles ($\theta_{12}, \theta_{13},
\theta_{23}$), and one complex phase ($\delta$); see
section~\ref{sec:notations} for definitions. We know that two out of the three
mixing angles are large~\cite{Schwetz:2008er},
\begin{equation}\label{eq:angles}
\sin^2\theta_{12} = 0.318^{+0.019}_{-0.016} \,,\qquad
\sin^2\theta_{23} = 0.50^{+0.07}_{-0.06} \,.
\end{equation}
The mass-squared differences are determined relatively accurately from the
spectral data in the KamLAND~\cite{:2008ee} and MINOS~\cite{Adamson:2008zt}
experiments, respectively~\cite{Schwetz:2008er},
\begin{equation}\label{eq:dmqs}
\Delta m^2_{21} = 7.59^{+0.23}_{-0.18} \times 10^{-5}~\eVq \,,\qquad
|\Delta m^2_{31}| = 2.40^{+0.12}_{-0.11} \times 10^{-3}~\eVq \,.
\end{equation}
The parameters in eqs.~\ref{eq:angles} and \ref{eq:dmqs} are responsible for
the dominating oscillation modes observed in the experiments mentioned
above. 

The topic of this review is the third mixing angle, $\theta_{13}$, whose
value is not known at present, and is constrained to be small compared to
the other two angles~\cite{Schwetz:2008er} (updated as of May 2010),
\begin{equation}
  \begin{array}{r@{\:\:\le\:\:}ll}
    \sin^2\theta_{13} & 0.031 & (0.047) \\
    \stheta           & 0.12 &(0.18)    \qquad\qquad
    90\%~(3\sigma)~\text{CL} \,.\\
    \theta_{13}       & 10.1^\circ & (12.5^\circ) 
  \end{array}
\end{equation}
An important contribution to the bound on $\theta_{13}$ comes from the
non-observation of disappearance of reactor electron anti-neutrinos at the
scale of $\dma$ at the CHOOZ~\cite{Apollonio:2002gd} and Palo
Verde~\cite{Boehm:2001ik} experiments, while the final bound is obtained
from the combination of global neutrino oscillation data, see
e.g.~\cite{Schwetz:2008er, Fogli:2008jx, GonzalezGarcia:2010er}. The present
information on $\theta_{13}$ is reviewed in some detail in
section~\ref{sec:present}, including also a discussion of possible hints for
a non-zero value~\cite{Fogli:2008jx, Collaboration:2009yc}.

Maybe besides the determination of the absolute neutrino mass and the search
for lepton number violation in neutrino-less double beta decay, the
determination of $\theta_{13}$ is one of the next primary goals in neutrino
physics. Its value is of great phenomenological as well as theoretical
interest. On the phenomenological side, the possibility of CP violation in
neutrino oscillations, which is a genuine three--flavour effect, depends on
a non-zero value of $\theta_{13}$. Any realistic possibility to determine
the type of the neutrino mass hierarchy (i.e., the sign of $\dma$) relies on
a not-too-small $\theta_{13}$. Therefore, the results on $\theta_{13}$ from
the upcoming generation of experiments will be of crucial importance for a
possible subsequent high-precision neutrino oscillation facility.  We
briefly comment on the issues of CP violation and mass hierarchy
determination in sections~\ref{sec:CP} and \ref{sec:MH}, respectively.
Implications of the value of $\theta_{13}$ for neutrino mass models and
flavour symmetries are briefly mentioned in section~\ref{sec:models}.

There are several neutrino oscillation experiments currently under
construction, which are expected to start data taking soon. These are the
reactor neutrino experiments Daya Bay~\cite{Guo:2007ug}, Double
Chooz~\cite{Ardellier:2006mn}, RENO~\cite{RENO-cdr} and the accelerator
experiments NO$\nu$A~\cite{Ambats:2004js} and T2K~\cite{Itow:2001ee}. The
primary goal for all of these experiments is the discovery of the yet
unknown mixing angle $\theta_{13}$.  Section~\ref{sec:exp} is devoted to a
description of these five experiments, and in section~\ref{sec:pheno} we
discuss the related phenomenology, highlighting the different nature of the
experiments ($\bar\nu_e$ disappearance in reactors versus $\nu_\mu\to\nu_e$
or $\bar\nu_\mu\to\bar\nu_e$ appearance in accelerators) and provide
sensitivity estimates.

The results of these experiments will be essential for the planning towards
a possible next generation of long-baseline neutrino experiments able to
address leptonic CP violation and the neutrino mass hierarchy. This could be
an upgraded super beam experiment with a huge detector, or experiments using
a new source for an intense neutrino beam based on decaying particles in a
storage ring, such as a neutrino factory~\cite{Geer:1997iz} (using muons) or
a beta beam~\cite{Zucchelli:2002sa, Volpe:2006in, Lindroos:2010zza} (using
radioactive ions). These options are under intense study, see
e.g.,~\cite{Bandyopadhyay:2007kx, Barger:2007yw, euronu, ids,
Bernabeu:2010rz} and any decisions will be crucially influenced by the
results of the upcoming experiments discussed in this review.  We briefly
discuss such future high precision facilities in section~\ref{sec:future}.

We focus in this paper on upcoming experiments which are under construction
or (from current perspective) are very likely to be funded. 
Without going into any details, let us just mention here also other more
speculative ideas towards a $\theta_{13}$ measurement. These include a
high-intensity tritium source immersed into a large
TPC~\cite{Giomataris:2003bp, Aune:2005is} or the use of M\"ossbauer
neutrinos~\cite{Raghavan:2005gn, Minakata:2006ne, Potzel:2009pr,
Kopp:2009fa, Akhmedov:2008jn}. Such approaches would lead to $\theta_{13}$
oscillations at baselines of order 10~m, thanks to the low energy of
neutrinos emitted in tritium decay ($Q = 18.6$~keV). Unfortunately serious
practical as well as principal problems seem to make such experiments
unlikely from present perspective.
The observation of neutrinos from a supernova explosion might also allow
some conclusions on $\theta_{13}$, in lucky circumstances even down to
values of order $10^{-5}$, see for example~\cite{Lunardini:2003eh, 
Gava:2009pj, Dasgupta:2010ae}.  Apart from the explosion of a nearby
supernova this requires one or even more suitable detectors at the correct
location on earth with respect to the direction of the supernova.
$\theta_{13}$ in the context of leptonic unitarity triangles has been
discussed in~\cite{Farzan:2002ct}, implications for neutrino-less
double beta
decay have been considered in~\cite{Lindner:2005kr}, discussions about short
and long baselines electron (anti)neutrino disappearance signals have been
published in \cite{Giunti:2009zz}.
Implications of $\theta_{13}$ for ultra-high energy neutrinos searched for
in neutrino telescopes have been discussed e.g., in~\cite{Serpico:2005sz,
Winter:2006ce}, and $\theta_{13}$ effects for neutrinos from WIMP
annihilations in the centre of the sun have been considered for example
in~\cite{Blennow:2007tw, Lehnert:2010vb}.

Throughout this paper we restrict ourselves to the simplest unitary
three--flavour framework, and ignore the possibility of additional sterile
neutrinos as well as other new physics such as non-standard neutrino
interactions. Let us note that effects of sterile neutrinos are strongly
constrained~\cite{Maltoni:2007zf, Karagiorgi:2009nb} and one does not expect
that they have an impact at the sensitivity level of the next generation of
experiments discussed here. This might be different for a subsequent
generation of high precision experiments.
While model independent bounds on non-standard neutrino interactions 
are at the level of $10^{-2}-10^{-1}$~\cite{Biggio:2009nt}, in
typical theories beyond the Standard Model one expects them to be much
smaller~\cite{Antusch:2008tz, Gavela:2008ra}, beyond the sensitivities of
the experiments considered here.

\section{General remarks on $\theta_{13}$ phenomenology}
\label{sec:general}

\subsection{Notations and conventions}
\label{sec:notations}

Three--flavour lepton mixing is described by a $3\times 3$ unitary matrix
$U$, the so-called PMNS mixing matrix~\cite{Maki:1962mu, Pontecorvo:1967fh}.
In a basis where the charged lepton mass matrix is diagonal we have
\begin{equation}
\nu_\alpha = \sum_{i=1}^{3} U_{\alpha i} \nu_i \,.
\end{equation}
Here, $\nu_\alpha$ ($\alpha = e,\mu,\tau$) are the (left-handed)
so-called ``flavour fields'' participating in charged current (CC)
interactions, and $\nu_i$ ($i=1,2,3$) are the neutrino mass eigenfields
with definite masses $m_1, m_2, m_3$, respectively. Neutrino
oscillations depend on the two independent mass-squared differences
$\Delta m^2_{21}$, $\Delta m^2_{31}$, with $\Delta m^2_{ij} \equiv
m^2_i - m_j^2$. After absorbing unphysical phases in $U$ by
redefining charged lepton fields, one finds that $U$ contains three
complex phases. Two of these are so-called Majorana phases, which
appear only in lepton number violating processes (such as for example
neutrino-less double beta decay) but drop out in neutrino oscillations
which depend only on the so-called Dirac phase. By convention $U$ is
parametrised by the three mixing angles
$\theta_{12},\theta_{23},\theta_{13}$, and the Dirac phase $\delta$ in
the following way:
\begin{eqnarray} \fl
U  =  
  \left(\begin{array}{ccc} 
    1&0&0\\ 0 & c_{23} & s_{23}\\ 0 & -s_{23} & c_{23}
  \end{array}\right) 
  \left(\begin{array}{ccc}
    c_{13}&0 & e^{-i\delta}s_{13} \\ 0 & 1 & 0 \\ 
    -e^{i\delta} s_{13}&0 & c_{13} 
  \end{array}\right)
  \left(\begin{array}{ccc}
    c_{12} & s_{12}& 0\\ -s_{12} & c_{12}&0\\0&0&1 
  \end{array}\right)\, D_\mathrm{Maj}  \label{eq:U1}\\
\fl\quad  =
  \left(\begin{array}{ccc}
  c_{12}c_{13} & 
  s_{12}c_{13} & s_{13}e^{-i\delta}\\
  -c_{23}s_{12}-s_{13}s_{23}c_{12}e^{i\delta} &
  c_{23}c_{12}-s_{13}s_{23}s_{12}e^{i\delta}  &
  s_{23}c_{13}\\
  s_{23}s_{12}-s_{13}c_{23}c_{12}e^{i\delta} &
  -s_{23}c_{12}-s_{13}c_{23}s_{12}e^{i\delta} &
  c_{23}c_{13}
  \end{array}
  \right) \, D_\mathrm{Maj} \, , \label{eq:U2}
\end{eqnarray}
with the abbreviations $s_{jk} \equiv \sin\theta_{jk}$, $c_{jk} \equiv
\cos\theta_{jk}$, and $D_\mathrm{Maj} \equiv
\text{diag}(e^{i\frac{\alpha}{2}}, e^{i\frac{\beta}{2}}, 1)$ with $\alpha$
and $\beta$ the Majorana phases. The ranges for the mixing angles and the
Dirac phase are \cite{Gluza:2001de} $0\le \theta_{jk} \le \pi/2$ and $0\le
\delta < 2\pi$. The evolution of the neutrino flavour state can be described
by a Schr\"odinger-like evolution equation with the Hamiltonian
\begin{equation}\label{eq:ham}
H_\nu = \frac{1}{2E_\nu} U \text{diag}(0, \dms, \dma) U^\dagger +
\text{diag}(V,0,0) \,,
\end{equation}
where $V$ is the effective potential in matter responsible for the MSW
matter effect~\cite{Wolfenstein:1978ue, Barger:1980tf, Mikheev:1985gs, Mikheev:1986wj},
with $V = \sqrt{2} G_F N_e$ where $N_e$ is the electron density along the
neutrino path. Eq.~\ref{eq:ham} holds for neutrinos, for anti-neutrinos one
has to replace $U \to U^*$ and $V\to -V$.
From eqs.~\ref{eq:U1}, \ref{eq:U2}, \ref{eq:ham} one can make the following
observations:
\begin{itemize}
\item
We see from eq.~\ref{eq:U2} that $|U_{e3}| = \sin\theta_{13}$. Therefore,
$\theta_{13}$ controls the fraction of the electron neutrino $\nu_e$
contained in the neutrino mass eigenfield corresponding to the mass $m_3$.
Since $\dms \ll |\dma|$ (see eq.~\ref{eq:dmqs}), $m_3$ corresponds to the
mass state separated by a larger gap from the other two mass states $m_1$
and $m_2$. Hence, for $\theta_{13} = 0$ the electron neutrino does not mix
with $\nu_3$ and can only participate in oscillations with $\dms$. This
means $\nu_e$ decouples from oscillations involving $m_3$, i.e.,
oscillations with $\dma \approx \Delta m^2_{32}$.
\item
Let us consider the limit $\theta_{13} \to 0$ and $\dms / |\dma| \to 0$. In
this case the evolution of solar neutrinos and oscillations in the KamLAND
reactor experiment are described by an effective two--flavour system of
$\nu_e$ and a combination of $\nu_\mu$ and $\nu_\tau$, governed by the
``solar parameters'' $\theta_{12}$ and $\dms$. Oscillations of atmospheric
neutrinos and in the K2K and MINOS long-baseline $\nu_\mu$ disappearance
experiments are described by two--flavour $\nu_\mu$-$\nu_\tau$ oscillations
governed by the ``atmospheric parameters'' $\theta_{23}$ and $\dma$. These
``atmospheric'' oscillations are pure vacuum oscillations, as it is easy to
see from eq.~\ref{eq:ham} that the matter potential decouples from the
relevant $2\times 2$ block of the evolution Hamiltonian in this limit.
\item
If $\theta_{13} > 0$ the electron neutrino participates also in oscillations
with $\dma$. Therefore, one can determine $\theta_{13}$ by looking for
transitions at the ``atmospheric'' scale involving the electron neutrino
flavour.
\end{itemize}

\subsection{Leptonic CP violation}
\label{sec:CP}

Leptogenesis~\cite{Fukugita:1986hr} provides a very attractive mechanism to
explain the generation of an matter--antimatter asymmetry in the early
Universe. An important ingredient for this to happen is CP violation in the
lepton sector. While in general there is no direct connection between the CP
violation at a very high scale necessary for leptogenesis and CP violation
observable in low-energy experiments, an observation of CP violation in
neutrino oscillations would provide a strong hint in favour of the
leptogenesis idea. Therefore, there is intense activity towards a
high-precision neutrino facility able to address CP violation in
oscillations~\cite{Bandyopadhyay:2007kx, Barger:2007yw, euronu, ids}.

CP violation in neutrino oscillations is a consequence of a non-trivial
Dirac phase $\delta$. From the parameterisation eq.~\ref{eq:U1} it is
obvious that $\delta$ becomes unphysical if $\theta_{13}$ is zero. Leptonic
CP violation will manifest itself in a difference of the vacuum oscillation
probabilities for neutrinos and anti-neutrinos~\cite{Cabibbo:1977nk,
Bilenky:1980cx, Barger:1980jm}
\begin{equation}
P_{\nu_\alpha \to \nu_\beta} - P_{\bar\nu_\alpha \to \bar\nu_\beta} = 
-16 \, J_{\alpha\beta} \sin\frac{\dms L}{4E_\nu}
\sin\frac{\Delta m^2_{32} L}{4E_\nu} \sin\frac{\dma L}{4E_\nu} \,,
\end{equation}
where
\begin{equation}
J_{\alpha\beta} = 
\text{Im}(U_{\alpha 1}U_{\alpha 2}^* U_{\beta 1}^* U_{\beta 2}) =
\pm J \,,\qquad J = s_{12} c_{12} s_{23} c_{23} s_{13} c_{13}^2 \sin\delta
\end{equation}
with $+(-)$ for (anti-)cyclic permutation of the indices $e,\mu,\tau$. $J$
is the leptonic analogue to the Jarlskog-invariant in the quark
sector~\cite{Jarlskog:1985ht}, which is a unique and parameterisation
independent measure for CP violation. It vanishes if any of the three mixing
angles is zero. Hence, a non-zero $\theta_{13}$ is a necessary prerequisite
for leptonic CP violation. A recent review on CP violation in neutrino
oscillations can be found here~\cite{Nunokawa:2007qh}.

\subsection{The neutrino mass hierarchy}
\label{sec:MH}

An important question in neutrino physics is the type of the neutrino mass
hierarchy, which can be normal (NH, $\dma > 0$) or inverted (IH, $\dma <
0$). This question has important consequences for possible neutrino mass
models, and the problem of flavour in general. Therefore, the determination
of the sign of $\dma$ is among the main goals for a future long-baseline
facility, see e.g.~\cite{Bandyopadhyay:2007kx}. Here we comment briefly on
the importance of the value of $\theta_{13}$ for this measurement.

The most promising way to distinguish these two neutrino mass spectra is to
search for the matter effect in transitions due to $\dma$. The condition for
an MSW resonance is 
\begin{equation}\label{eq:res}
\cos 2\theta_{13} = \pm \frac{2 E_\nu V}{\dma} \,,
\end{equation}
where $+(-)$ holds for (anti)neutrinos. For a given sign of $\dma$,
eq.~\ref{eq:res} can be fulfilled either for neutrinos or for anti-neutrinos.
Therefore, finding out whether the matter resonance due to $\dma$ occurs for
neutrinos or anti-neutrinos will provide a determination of the sign of
$\dma$. As discussed above the occurrence of a matter effect in $\dma$
transitions is only possible for a non-zero $\theta_{13}$, and hence the
possibility to determine the neutrino mass hierarchy via the matter effect
crucially depends on the observability of $\theta_{13}$ effects. This can be
done in long-baseline experiments~\cite{Freund:1999gy, Barger:2000cp} or with
atmospheric neutrinos~\cite{Bernabeu:2003yp, Indumathi:2004kd, Huber:2005ep,
Petcov:2005rv}.

In principle changing the sign of $\dma$ has also implications for 
disappearance oscillation probabilities in vacuum, independent of the matter
effect~\cite{Petcov:2001sy, Nunokawa:2005nx, deGouvea:2005hk}. While this
effect for muon neutrino disappearance is independent of $\theta_{13}$, in
the case of electron neutrino disappearance also these methods require a
large value of $\theta_{13}$. In practice, however, such measurements turn
out to be extremely difficult, maybe unrealistic~\cite{deGouvea:2005mi,
Gandhi:2009kz}. For completeness we also mention the possibility to
discriminate NH and IH in non-oscillation experiments,
e.g.~\cite{Pascoli:2005zb, Choubey:2005rq}, or from supernova neutrino
observations, e.g.~\cite{Lunardini:2003eh, Dighe:2003be, Duan:2007bt,
Dasgupta:2008my, Dasgupta:2010cd}.

\subsection{$\theta_{13}$ and models for neutrino mass}
\label{sec:models}

On the theoretical side, the determination of $\theta_{13}$ will provide
important information on the mechanism of neutrino mass generation and the
flavour structure in the lepton sector. Naively one may expect that since
two mixing angles are large also the third one should not be too small.
Considering neutrino mass models without any flavour structure, so-called
anarchical models, one does expect a value of $\theta_{13}$ close to the
present bound~\cite{deGouvea:2003xe}. If on the contrary experiments would
indicate a very tiny value for $\theta_{13}$ one might wish to have a
symmetry reason as an explanation. 

The intriguing result that mixing in the lepton sector is very
different from the quark sector might indicate that a special
mechanism is at work to produce the peculiar values of the lepton
mixing angles.
For example, rather symmetric patterns for the mixing matrix are the
tri-bimaximal~\cite{Harrison:2002er} or the
bimaximal~\cite{Barger:1998ta} mixing matrices,
\begin{equation}\fl
U_\text{tri-bimax} = 
\left(\begin{array}{ccc}
\!\sqrt{2/3}&1/\sqrt{3} &0\!\\
\!-1/\sqrt{6}&1/\sqrt{3}&1/\sqrt{2}\!\\
\!1/\sqrt{6}&-1/\sqrt{3}&1/\sqrt{2}\!
\end{array}
\right) \,,\quad
U_\text{bimax} = 
\left(\begin{array}{ccc}
\!1/\sqrt{2}& 1/\sqrt{2}&0\!\\
\!-1/2&1/2&1/\sqrt{2}\!\\
\!1/2&-1/2&1/\sqrt{2}\!
\end{array}
\right) \,.
\label{eq:tri-bi}
\end{equation}
In both cases $s_{23}^2 = 1/2$ (maximal 2-3 mixing) and
$\theta_{13}=0$.  While for tri-bimaximal mixing $s_{12}^2 = 1/3$ (in
perfect agreement with data, see eq.~\ref{eq:angles}), for bimaximal
mixing $s_{12}^2 = 1/2$, which requires significant corrections to be
consistent with the experimental value. 

The measured values of the mixing angles may be the result of such
regular patterns. This could indicate a special symmetry among
generations. There is a huge literature on flavour symmetries, which
we cannot properly account for here.  Popular symmetries are for
example a $\mu$-$\tau$ exchange symmetry~\cite{Lam:2001fb,
Grimus:2003vx} (note that both examples given in eq.~\ref{eq:tri-bi}
fulfil the relation $|U_{\mu i}| = |U_{\tau i}|$) or the $A_4$
permutation symmetry~\cite{Ma:2001dn} (usually in context of
tri-bimaximal mixing), see~\cite{Altarelli:2010gt} for a recent review
and references.
In many cases flavour symmetries predict $\theta_{13} = 0$ in the
limit where the symmetry is exact, as in the tri-bimaximal and
bimaximal examples given above. However, one expects corrections to
the zeroth order symmetric limit due to breaking of the symmetry,
which then in general will induce a finite value for $\theta_{13}$,
see for example~\cite{Hochmuth:2007wq, Ge:2010js}.  Furthermore, if
the symmetry holds at some high scale, renormalisation group running
will introduce deviations from the symmetric limit, see
e.g.~\cite{Antusch:2005gp}. In~\cite{King:2009qt} a model is discussed
which allows for a quite sizeable value of $\theta_{13}$ but still
preserves the tri-bimaximal predictions for $\theta_{12}$ and
$\theta_{23}$.

An alternative approach to flavour comes from Grand Unified Theories (GUTs).
A priori a GUT by it self makes no statement about flavour (unless it is
complemented with additional flavour symmetries). However, since quarks and
leptons reside in common representations of the GUT gauge group, one obtains
relations between quark and lepton masses and mixing parameters. For
example, a specific model based on the SO(10) gauge group has been discussed
in~\cite{Bertolini:2006pe}, where the predictions for $\theta_{13}$ based on
GUT relations between the Yukawa matrices have been worked out explicitly.
Predictions of this kind can be confronted with future measurements of
$\theta_{13}$. 

In Ref.~\cite{Albright:2009cn} a survey of many neutrino mass models has
been performed, with a particular focus on the predictions for
$\theta_{13}$, see also table~1 of~\cite{Anderson:2004pk}. Here we just want
to point out the importance of the $\theta_{13}$ measurement for
discriminating among models. Certainly an improved determination of
$\theta_{13}$ (either establishing a finite value or setting a more
stringent upper bound) will provide important new information regarding the
problem of flavour. Further discussion and references can be found in the
review~\cite{Mohapatra:2005wg}.

\section{Present status and possible hints for $\theta_{13} > 0$}
\label{sec:present}

\begin{figure}
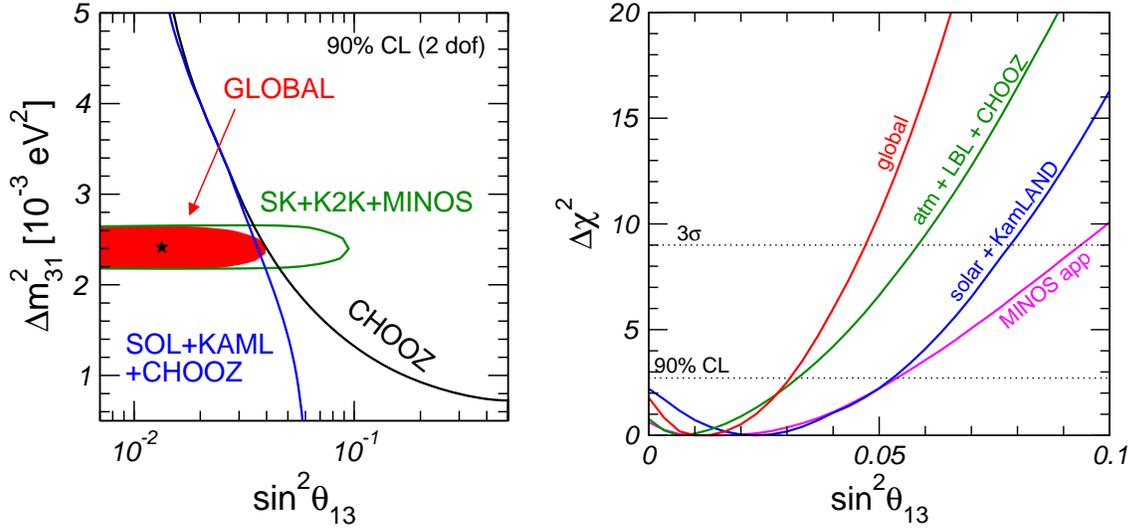

  \centering
  \includegraphics[height=7cm]{f.1a.th13-2010-lin.eps}
  \quad
  \includegraphics[height=7cm]{f.1b.F-th13-2010-tot-lin.eps}
  \caption{Constraints on $\sin^2\theta_{13}$ from different parts of
  the global data~\cite{Schwetz:2008er} (updated as of May 2010).}
  \label{fig:th13-bound}
\end{figure}

The present information on the value of $\theta_{13}$ emerges from an
interplay of the global data on neutrino oscillations, as illustrated in 
fig.~\ref{fig:th13-bound}, see~\cite{Schwetz:2008er, GonzalezGarcia:2010er,
Fogli:2009ce} for recent global analyses. In the following we discuss the
phenomenology of the various different data sets from the CHOOZ reactor
experiment (sec.~\ref{sec:chooz}), solar neutrinos and the KamLAND reactor
experiment (sec.~\ref{sec:sol}), atmospheric neutrinos (sec.~\ref{sec:atm}),
and the $\nu_e$ appearance data from the MINOS long-baseline experiment
(sec.~\ref{sec:minos}).  In section~\ref{sec:global} we summarise the status
of $\theta_{13}$ in the global analysis.

\subsection{The bound from the CHOOZ reactor experiment}
\label{sec:chooz}

An important contribution to the bound comes, of course, from the
CHOOZ reactor experiment~\cite{Apollonio:2002gd}. Experimental aspects
and the phenomenology of $\theta_{13}$ reactor experiments will be
discussed below in sections~\ref{sec:react} and \ref{sec:pheno}.  The
CHOOZ experiment observed the $\bar\nu_e$ flux emitted from the two
cores of the Chooz nuclear power plant at a distance $L \approx
1.05$~km. The survival probability is given by
\begin{equation}
P_{ee} \approx 1 - \stheta \sin^2\frac{\dma L}{4 E_\nu} +
\mathcal{O} (\alpha^2)\,,
\end{equation}
with $\alpha \equiv \dms/\dma$. Considering typical reactor neutrino
energies of $E_\nu \sim 4$~MeV one finds that this experiment is sensitive
to the $\theta_{13}$ induced oscillations with $\dma$, whereas terms due to
$\dms$ are suppressed by the small solar mass squared difference and can be
neglected as long as $\stheta \gtrsim \alpha^2 \simeq 10^{-3}$. The ratio of
the measured and the predicted neutrino rates in CHOOZ, averaged over the
energy spectrum, has been obtained as~\cite{Apollonio:2002gd}
\begin{equation}
R = 1.01 \pm 2.8\% (\text{stat}) \pm 2.7\% (\text{syst}) \,.
\label{eq:Chooz-result}
\end{equation}
The parameter constraint resulting from this measurement is shown in
fig.~\ref{fig:th13-bound} in the plane of $\sin^2\theta_{13}$ and $\dma$.
A less constraining result was reported also by the Palo Verde
reactor experiment \cite{Boehm:2001ik}.
A meaning full bound on $\theta_{13}$ can only be obtained by combining the
CHOOZ result with the determination of $|\dma|$ from atmospheric and
long-baseline experiments. From the combined analysis of CHOOZ + atmospheric
neutrino data + K2K + MINOS disappearance data one finds $\sin^2\theta_{13} <
0.027$ at 90\%~CL~\cite{Schwetz:2008er}.

\subsection{Information from solar neutrinos and the KamLAND reactor experiment}
\label{sec:sol}

For solar neutrinos as well as the KamLAND reactor experiment it is an
excellent approximation to consider the limit $\dma \to \infty$. Then
it follows from the Hamiltonian in eq.~\ref{eq:ham} with the
parametrisation for $U$ from eq.~\ref{eq:U1}, that the survival
probability of electron (anti)neutrinos is given by \cite{Kuo:1986sk,
Shi:1991zw, Goswami:2004cn}
\begin{equation}
P_{ee} \approx c_{13}^4 P_{ee}^{2\nu} + s_{13}^4 \,,
\end{equation}
where $P_{ee}^{2\nu}$ is a two--flavour survival probability depending on
$\theta_{12}$ and $\dms$, where the matter potential $V$ is replaced by
$c_{13}^2 V$. The $s_{13}^4$ term is tiny and will be neglected in the
following discussion.

For the KamLAND reactor experiment the matter potential can be neglected and
$P_{ee}^{2\nu}$ is just the vacuum probability:
\begin{equation}
P_{ee}^\mathrm{KamL} \approx
\cos^4\theta_{13} \left(1-\sin^22\theta_{12}\sin^2\frac{\dms L}{4E_\nu}\right) \,.
\end{equation}
This leads to an anti-correlation of $\sin^2\theta_{13}$ and
$\sin^2\theta_{12}$~\cite{Maltoni:2004ei}, see
also~\cite{Goswami:2004cn, Balantekin:2008zm}. For solar neutrinos one
obtains simple expressions for the low and high energy part of the
solar neutrino spectrum, below and above the MSW resonance in the sun
at $E_\nu^\mathrm{res} \approx 2$~MeV. For low energy solar neutrinos
one can adopt the vacuum approximation, and averaging over the fast
oscillations leads to
\begin{equation}\label{eq:sun-low}
P_{ee}^\mathrm{solar} \approx
\cos^4\theta_{13} \left(1- \frac{1}{2}\sin^22\theta_{12}\right) \qquad
\mbox{(low energies)} \,. 
\end{equation}
The high energy part of the spectrum undergoes the adiabatic MSW conversion
inside the sun and $P_{ee}^{2\nu} \approx \sin^2\theta_{12}$:
\begin{equation}\label{eq:sun-high}
P_{ee}^\mathrm{solar} \approx
\cos^4\theta_{13} \, \sin^2\theta_{12} \qquad
\mbox{(high energies)} \,.
\end{equation}
Eq.~\ref{eq:sun-low} for low energy solar neutrinos shows a similar
anti-correlation between $\sin^2\theta_{13}$ and $\sin^2\theta_{12}$ as in
KamLAND, whereas for the high energy flux subject to the SNO CC/NC
measurement~\cite{Aharmim:2008kc, Collaboration:2009gd}, a positive
correlation of $\sin^2\theta_{13}$ and $\sin^2\theta_{12}$ emerges. As
discussed e.g.~in~\cite{Maltoni:2004ei, Goswami:2004cn}, this
complementarity leads to a non-trivial constraint on $\theta_{13}$.

\begin{figure}
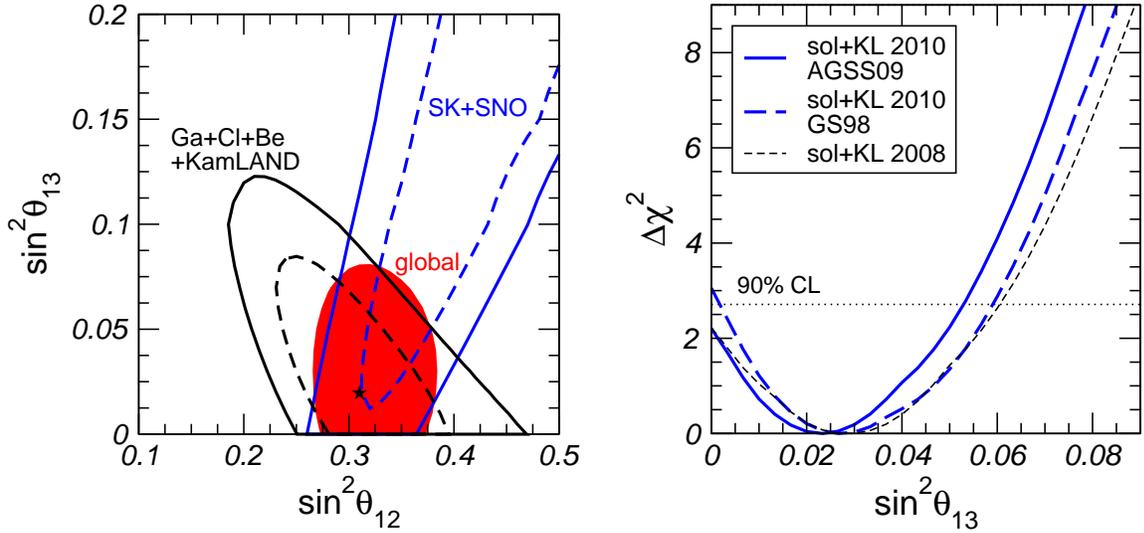

  \centering
 \includegraphics[height=0.45\textwidth]{f.2a.s12-s13-HE-LE.eps}
 \quad
 \includegraphics[height=0.45\textwidth]{f.2b.F-th13-2010-solar.eps}
  \caption{Left: Allowed regions in the $(\theta_{12}-\theta_{13})$ plane at
  90\% and 99.73\%~CL (2~dof). The black contours correspond to KamLAND
  combined with the low energy solar neutrino experiments Homestake, the
  Gallium experiments, and Borexino, whereas the blue contours correspond to
  the high energy solar neutrino experiments SuperK and SNO. The red shaded
  region shows the 99.73\%~CL region for the combined analysis. $\dms$ has
  been fixed at the combined best fit point. (Figure kindly produced by
  Mariam Tortola.) Right: $\Delta\chi^2$ as a function of
  $\sin^2\theta_{13}$ for solar and KamLAND data, illustrating the impact of
  the most recent solar neutrino data (mainly the SNO low energy threshold
  analysis~\cite{Collaboration:2009gd}) as well as the impact of different
  solar models (AGSS09 and GS98 refer to low and high metalicity solar
  models, respectively~\cite{Serenelli:2009yc}). \label{fig:th13-sol-kl}}
\end{figure}

Here we present results from the global solar neutrino and KamLAND analysis
of Ref.~\cite{Schwetz:2008er} (2010 updated arXiv version~3), where a
description of the used data can be found. For technical details see also
\cite{Maltoni:2004ei, Maltoni:2003da}. Fig.~\ref{fig:th13-sol-kl}~(left)
illustrates the different correlations between $\theta_{12}$ and
$\theta_{13}$ from KamLAND/low energy solar and high energy solar data
discussed above. The right panel shows the $\Delta\chi^2$ profiles for two
different assumptions on the Standard Solar Model. The curves labeled AGSS09
and GS98 refer to low and high metalicity solar models,
respectively~\cite{Serenelli:2009yc}, which imply slightly different
predictions for the $^8$B solar neutrino flux.  The ``default'' assumption
in the global analysis of~\cite{Schwetz:2008er} is the AGSS09 low metalicity
solar model. For a discussion of the impact of solar models on the fit see
also~\cite{GonzalezGarcia:2010er}. 
Furthermore, the thin black dashed curves shows the $\Delta\chi^2$ profile
for the 2008 solar neutrino analysis from~\cite{Schwetz:2008er},
highlighting the impact of the most recent solar neutrino data, mainly the
SNO low energy threshold analysis~\cite{Collaboration:2009gd} which leads to
an improved determination of the total NC event rate with the remarkable
precision of 3.1\% (stat) and 2.5\% (syst) at $1\sigma$.

The $\theta_{13}$ fit results~\cite{Schwetz:2008er} from global solar
data and KamLAND are (best fit and $1\sigma$ errors)
\begin{equation}\label{eq:th13-sol-kl}
\sin^2\theta_{13} = 0.022^{+0.018}_{-0.015} \qquad \text{(solar + KamLAND)} \,,
\end{equation}
and the following bound is obtained at 90\% ($3\sigma$)~CL:
\begin{equation}
\sin^2\theta_{13} < 0.053~(0.078) \qquad \text{(solar + KamLAND)} \,.
\end{equation}
The result from eq.~\ref{eq:th13-sol-kl} is in excellent agreement
with the value obtained by the SNO
Collaboration~\cite{Collaboration:2009gd}, $\sin^2\theta_{13} =
0.0200^{+0.0209}_{-0.0163}$, and it implies a slight preference for a
non-zero value of $\theta_{13}$. At $\theta_{13}=0$ we have $\Delta
\chi^2 = 2.2$ (see fig.~\ref{fig:th13-sol-kl}), which corresponds to a
hint at about $1.5\sigma$. This hint emerges from a slight mismatch of
the best fit points for $\theta_{12}$ and $\dms$ for $\theta_{13} = 0$
for solar and KamLAND data separately~\cite{Fogli:2008jx,
Goswami:2004cn, Balantekin:2008zm, Maltoni:2003da}. This mismatch can
be alleviated to some extent by a finite value of $\theta_{13}$, see
e.g.~\cite{Balantekin:2008zm} for a discussion. From
fig.~\ref{fig:th13-sol-kl} (left) we see that the high energy solar
neutrino experiments SuperK and SNO show a slight preference for a
non-zero $\theta_{13}$. Eq.~\ref{eq:sun-high} indicates a degeneracy
among $\theta_{13}$ and $\theta_{12}$ as visible in the figure. This
degeneracy is not complete due to additional information from the
spectral shape as well as day-night asymmetries induced by the matter
effect in the earth during night. While the $\chi^2$ slope along the
SuperK+SNO allowed region is very flat, these data pull the global
best fit point somewhat towards a non-zero value.
For models with higher solar metallicities like GS98, a slightly larger best
fit point is obtained, $\sin^2\theta_{13} = 0.027^{+0.019}_{-0.015}$ and
$\Delta \chi^2 = 3.05$ at $\theta_{13} = 0$, while the bound on
$\theta_{13}$ becomes slightly weaker, see fig.~\ref{fig:th13-sol-kl}
(right).

\subsection{$\theta_{13}$ and atmospheric neutrinos}
\label{sec:atm}

Atmospheric neutrino data come from the SuperKamiokande experiment
that recently released an updated analysis
of the combined SK I+II+II data~\cite{Wendell:2010md}.
Atmospheric neutrinos provide a probe of several decades in $L/E_\nu$
thanks to the wide range in neutrino energies, from few 100~MeV to
several GeV, and baselines of $L\sim 15$~km for down going neutrinos
up to $L \sim 12\,000$~km for up-going ones. Furthermore, up-going
neutrinos have large trajectories through earth matter, opening the
possibility to look for matter effects. On the other hand, the
atmospheric neutrino ``beam'' is very messy, consisting of $\nu_\mu,
\bar\nu_\mu, \nu_e, \bar\nu_e$ fluxes with large
uncertainties. Therefore, from the observation of muons and electrons
from neutrino CC interactions one can access only certain combinations
of transition and survival probabilities. Also, the neutrino direction
for a given CC event is not known.

In water \ceren detectors such as SuperKamiokande, electron-like events
provide sensitivity to three--flavour effects. One can identify three type
of effects: $\theta_{13}$-induced~\cite{Petcov:1998su, Akhmedov:1998ui,
Akhmedov:1998xq, Chizhov:1998ug, Chizhov:1999az, Chizhov:1999he,
Bernabeu:2003yp, Akhmedov:2006hb}, $\dms$-induced~\cite{Kim:1998bv,
Peres:1999yi, Peres:2003wd, Gonzalez-Garcia:2004cu, Akhmedov:2008qt}, and
interference effects of the previous two~\cite{Peres:2003wd}, see
e.g.~\cite{Fogli:2005cq} for a discussion.  Effects induced by $\theta_{13}$
are important in the multi-GeV energy range, whereas $\dms$ effects are
mainly relevant for sub-GeV energies. Defining the excess of $e$-like events
as $\epsilon_e \equiv (N_e/N_e^0 - 1)$, with $N_e \, (N_e^0)$ being the
number of $e$-like events with (without) oscillations, one has
\begin{eqnarray}
\epsilon_e^\mathrm{multi} &\approx
\left( r \, \sin^2\theta_{23} - 1 \right) 
\langle P_{31}^{2\nu} \rangle \,,
\label{eq:excess-multi}\\
\epsilon_e^\mathrm{sub} &\approx
\left( r \, \cos^2\theta_{23} - 1 \right) 
\langle P_{21}^{2\nu} \rangle \,.
\label{eq:excess-sub}
\end{eqnarray}
Here  $\langle P_{31}^{2\nu}\rangle$ ($\langle
P_{21}^{2\nu}\rangle$) is an effective two--flavour probability governed by
$\dma$ and $\theta_{13}$ ($\dms$ and $\theta_{12}$), appropriately averaged
and including the weighted contributions from neutrinos and
anti-neutrinos, and $r(E_\nu,\Theta) \equiv F_\mu^0 / F_e^0$ is the ratio of
the initial unoscillated muon and electron neutrino fluxes, $\Theta$ being
the zenith angle. 
Since for sub-GeV energies $r\approx 2$, $\dms$ effects are suppressed for
$\theta_{23} \approx \pi/4$, however they provide a sensitive measure for
deviations from maximal $\theta_{23}$ mixing. Indeed, small deviations from
$\theta_{23} = \pi/4$ found in~\cite{Fogli:2005cq, GonzalezGarcia:2007ib}
may be traced back to a slight excess of sub-GeV $e$-like events in the
SuperKamiokande data, see, however~\cite{Wendell:2010md}, where such
deviations are not found.

Here we are more interested in $\theta_{13}$ induced effects from
eq.~\ref{eq:excess-multi}, relevant in the multi-GeV range, where one
has $r \approx 2.6-4.5$. Hence one expects that these effects could
show up also for maximal $\theta_{23}$ mixing.  Qualitatively,
$\epsilon_e^\mathrm{multi}$ vanishes for $\theta_{13} = 0$ and
increases monotonically with $\theta_{13}$. The effect is most
pronounced for zenith angles corresponding to neutrino trajectories
crossing the earth mantle, or earth mantle and core, where
$\theta_{13}$-effects can be resonantly enhanced due to matter
effects~\cite{Wolfenstein:1978ue, Mikheev:1985gs, Petcov:1998su,
Akhmedov:1998ui}. In the relevant zenith angle bins
$\epsilon_e^\mathrm{multi}$ can reach values of the order of 10\% (see
e.g.\ fig.~5 of~\cite{Bernabeu:2003yp}). For the normal hierarchy the
resonant matter enhancement occurs for neutrinos, whereas for the
inverted hierarchy it occurs for anti-neutrinos. Since the event
numbers in water \ceren detectors are dominated by neutrinos because
of larger cross sections, $\epsilon_e^\mathrm{multi}$ is larger by a
factor of $1.5 - 2$ for the normal hierarchy than for the inverted
one.

In addition to $\dms$-effects of eq.~\ref{eq:excess-sub} and
$\theta_{13}$-effects of eq.~\ref{eq:excess-multi} also an interference term
between the two contributions is present~\cite{Peres:2003wd}. It is
proportional to $(r \, \sin\theta_{13} \, \sin 2\theta_{23})$, it vanishes
in the limit $\dms =0$, and it depends on the CP-phase $\delta$. Because
of the different dependence on the flux ratio $r$ the interference term may
become important in cases where the effects governed by
eqs.~\ref{eq:excess-sub} and \ref{eq:excess-multi} are
suppressed~\cite{Fogli:2005cq}.

Such three--flavour effects in atmospheric neutrinos can be used to
determine the neutrino mass hierarchy or the octant of $\theta_{23}$
(if different from $\pi/4$), in huge future atmospheric neutrino
detectors (e.g., water \ceren or magnetised iron), possibly in
combination with a future long-baseline experiment, see for
example~\cite{Indumathi:2004kd, Huber:2005ep, Petcov:2005rv,
TabarellideFatis:2002ni, Campagne:2006yx, Gandhi:2007td}. In the
following we discuss possible implications of present data from
SuperKamiokande (SK) on $\theta_{13}$.

In Ref.~\cite{Fogli:2005cq} a preference for a non-zero $\theta_{13}$ value
from SK-I data was noted (see also~\cite{Escamilla:2008vq}).
Refs.~\cite{Fogli:2005cq, Fogli:2008jx, Fogli:2009ce} find from atmospheric
+ CHOOZ + long-baseline disappearance data a $0.9\sigma$ hint for a non-zero
value: $\sin^2\theta_{13} = 0.012 \pm 0.013$. In the atmospheric neutrino
analysis in \cite{Schwetz:2008er} which is based on \cite{Maltoni:2004ei}
$\dms$ effects are neglected, and combined with CHOOZ data the best fit
occurs for $\theta_{13} = 0$ (c.f.\ fig.~\ref{fig:th13-minos}, right). This
is in agreement with a similar analysis by the SuperKamiokande
collaboration~\cite{Hosaka:2006zd, Wendell:2010md}. Also, in the atmospheric
neutrino analysis from Ref.~\cite{GonzalezGarcia:2007ib} (which does include
$\dms$ effects, as Refs.~\cite{Fogli:2005cq, Fogli:2008jx}) the preference
for a non-zero $\theta_{13}$ is much weaker than the one
from~\cite{Fogli:2005cq}, with a $\Delta\chi^2 \lesssim 0.2$. 

A discussion and comparison of the results of different groups can be found
in~\cite{Schwetz:2006dh}.
The possible origin of the hint from atmospheric data has been
investigated in some detail in~\cite{Maltoni:2008ka,
GonzalezGarcia:2010er}.  Ref.~\cite{Maltoni:2008ka} concludes that the
statistical relevance of the hint for non-zero $\theta_{13}$ from
atmospheric data depends strongly on the details of the event rate
calculations and of the $\chi^2$ analysis (such as e.g., the specific
way of how systematic uncertainties are treated). Furthermore, the
reason for the hint seems be a small excess of multi-GeV $e$-like
events in SK-I data (used in \cite{Fogli:2005cq, Fogli:2008jx}), which
however disappears after the inclusion of SK-II data.
Ref.~\cite{GonzalezGarcia:2010er} comes to a similar conclusion from
the combined SK-I+II+III data.

Let us stress, however, that all analyses agree within
$\Delta\chi^2\approx 1$ and therefore there is no significant
disagreement. Certainly these are very subtle effects sensitive to the
fine details of the data analysis which can be addressed optimally
only within the experimental collaboration.  The recent three--flavour
analysis by SuperKamiokande~\cite{Wendell:2010md} did not find any
hint for a non-zero $\theta_{13}$. The $\theta_{13}$ analysis of
\cite{Wendell:2010md} has been performed adopting the approximation
$\dms = 0$, and therefore, a direct comparison with the results of the
phenomenological groups~\cite{Fogli:2005cq, Fogli:2008jx,
Maltoni:2008ka, GonzalezGarcia:2007ib, GonzalezGarcia:2010er} is still
not possible.

\subsection{$\nu_e$ appearance data at accelerators}
\label{sec:minos}

The first result on \nue appearance at a long baseline experiment has
been published by the K2K collaboration \cite{Ahn:2004te}:
$\stheta \ge 0.3$ (90\% CL) in a two-flavour approximation analysis.
Now also the MINOS experiment provides a first glance at a $\nu_\mu\to\nu_e$
appearance search for $\theta_{13}$~\cite{Collaboration:2009yc}, which is
actually the main objective for the upcoming T2K and NO$\nu$A experiments,
see section~\ref{sec:exp}.

In Ref.~\cite{Collaboration:2009yc} a search for $\nu_\mu\to\nu_e$
transitions by the MINOS experiment has been presented, based on a
$3.14 \times 10^{20}$ protons-on-target (pot) exposure in the Fermilab NuMI
beam (out of the $7\times 10^{20}$ pot accumulated so far).
35 events have been observed in the far detector with a
background of $27 \pm 5{\rm (stat)}\pm 2 {\rm (syst)}$ events
predicted by the measurements in the near detector. This corresponds
to an excess of about $1.5\sigma$ which can be interpreted as a weak
hint for $\nu_e$ appearance due to a non-zero $\theta_{13}$.
\footnote{We note however that the signal excess depends very
much on the choice of the position of the cut on the neural
network variable that separates signal from backgrounds
(fig.~2 of reference~\cite{Collaboration:2009yc}).
Also for this reason the publication of the \nue appearance 
result with the full statistics is quite important.}

In the MINOS detector, being optimised for muons, it is rather
difficult to identify $\nu_e$~CC events since they lead to an
electromagnetic shower not very different from a $\pi^\circ$ signal.
Neutral current (NC) and misidentified
$\nu_\mu$~CC events often have a similar signature, and hence lead to
a background for the $\nu_e$ appearance search. Indeed, in
Ref.~\cite{Adamson:2008jh} an analysis of ``NC events'' has been
performed, where ``NC events'' in fact include also $\nu_e$~CC events
due to the similar event topology. Therefore, a possible $\nu_\mu
\to\nu_e$ oscillation signal would contribute to the ``NC event''
sample of~\cite{Adamson:2008jh} and these data can be used to
constraint $\theta_{13}$.

We have performed a fit to the MINOS $\nu_e$ appearance and NC data by
using the GLoBES simulation software~\cite{Huber:2004ka,
Huber:2007ji}. The predicted $\nu_e$ appearance spectrum has been
calibrated by using the information given in \cite{Boehm:2009zz}. In
the fit we include a 7.3\% uncertainty on the background normalisation
(Tab.~I of~\cite{Collaboration:2009yc}). For the NC analysis we have
performed a fit to the observed spectrum by summing the NC events
induced from the total neutrino flux with the $\nu_e$~CC appearance
signal due to oscillations. We include a 4\% error on the predicted NC
spectrum and a 3\% error on the $\nu_\mu$ CC induced background
(Tab.~II of~\cite{Adamson:2008jh}). A full three--flavour fit is
performed taking into account a 5\% uncertainty on the matter density
along the neutrino path.

\begin{figure}
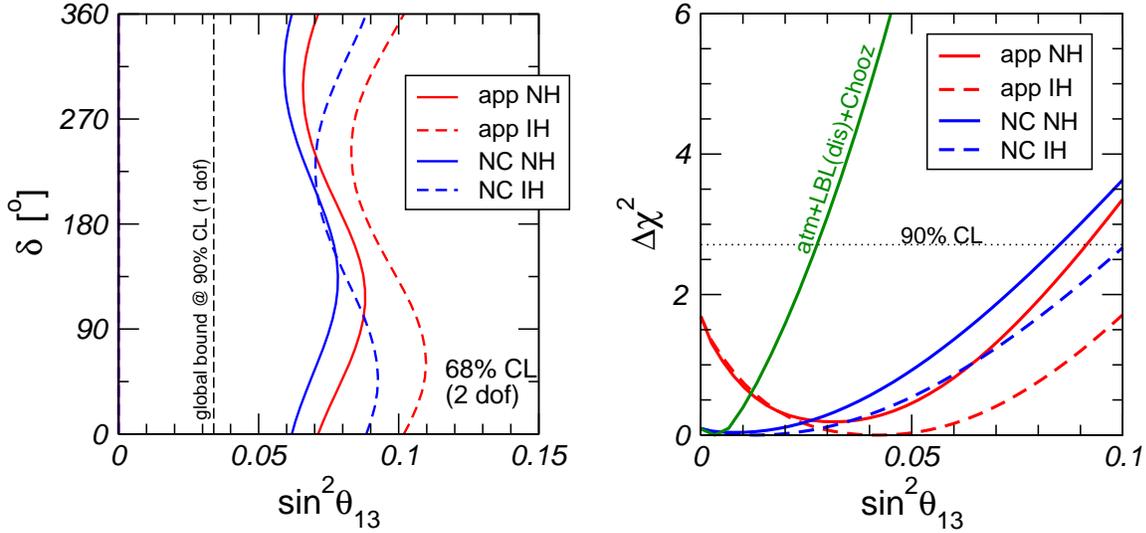

  \centering
 \includegraphics[height=0.45\textwidth]{f.3a.minos-del-th13.eps}
 \quad
 \includegraphics[height=0.45\textwidth]{f.3b.F-th13-2010-minos.eps}
  \caption{Left: Allowed regions in the $(\sin^2\theta_{13} - \delta)$ plane
  at 68\%~CL (2~dof) for MINOS $\nu_e$ appearance and NC data. Regions are
  shown separately for normal (NH) and inverted (IH) neutrino mass
  hierarchy. For comparison we show also the bound from global data at
  90\%~CL (1~dof).  Right: $\Delta\chi^2$ projection as a function of
  $\sin^2\theta_{13}$ for MINOS $\nu_e$ appearance and NC data, assuming NH
  (solid) and IH (dashed), both with respect to the common minimum, which
  occurs for IH. The green solid curve corresponds to the bound from CHOOZ +
  atmospheric + K2K + MINOS~(disappearance) data.\label{fig:th13-minos}}
\end{figure}

The $\nu_e\to\nu_\mu$ oscillation probability depends in a non-trivial way
on all 6 oscillation parameters, in particular on the unknown value of the
CP-phase $\delta$ as well as on the sign of $\dma$ (neutrino mass
hierarchy). The dependence has to be taken into account when extracting
information on $\theta_{13}$. This will be discussed in more detail in
section~\ref{sec:pheno} in the context of the upcoming long-baseline
accelerator experiments. 
Fig.~\ref{fig:th13-minos} shows the fit results for MINOS $\nu_e$
appearance and NC data, illustrating the importance of the parameter
correlations. The left panel shows the allowed regions in the
$(\sin^2\theta_{13} - \delta)$ plane, for both neutrino mass
hierarchies separately. We observe the typical S-shaped regions coming
from the trigonometric dependence on the phase $\delta$. In the right
panel the $\chi^2$ is marginalised with respect to all parameters
except for $\theta_{13}$ and sgn($\dma$), where for the solar and
atmospheric parameters we imposed Gaussian errors taken from
eqs.~\ref{eq:angles}, \ref{eq:dmqs}, without including any other
information on $\theta_{13}$ except from MINOS itself.

First, we note that the best fit point is always obtained for the
inverted hierarchy ($\Delta m^2_{31} < 0$), and in that case in
general the constraint on $\sin^2\theta_{13}$ is weaker, since for IH
the matter effect tends to suppress the $\nu_e$ appearance
probability. In fig.~\ref{fig:th13-minos} (right) the $\Delta\chi^2$
for normal hierarchy ($\Delta m^2_{31} > 0$) is given with respect to
the best fit for IH. In the global analysis we also marginalise over
the two hierarchies, and hence, the actual information from MINOS
comes from the IH.

Second, we see from the figure that MINOS $\nu_e$ appearance data
shows a slight preference for a non-zero value of $\theta_{13}$, with
a best fit point of $\sin^2\theta_{13} = 0.032(0.043)$ for NH (IH)
with $\Delta\chi^2 = 1.8$ at $\sin^2\theta_{13} = 0$ (about
$1.3\sigma$). In contrast, no indication for a non-zero $\theta_{13}$
comes from the NC data. Furthermore, one observes that NC gives a
slightly more constraining upper bound on $\sin^2\theta_{13}$ than
$\nu_e$ appearance, while both are significantly weaker than the bound
from $\nu_\mu$ disappearance data + CHOOZ. Let us mention that the
result for the NC analysis strongly depends on the value assumed for
the systematic uncertainty, whereas the $\nu_e$ appearance result is
more robust with respect to systematics, being dominated by
statistics.

MINOS $\nu_e$ and NC data are not independent, and adding the
corresponding $\chi^2$'s would imply a double counting of the same
data. Therefore, only $\nu_e$ appearance data without the information
from NC data is used in the global analysis
of~\cite{Schwetz:2008er}. It is found, however, that adding both MINOS
data sets (ignoring the double counting problem) leads to practically
the same result in the global fit, both for the ``hint'' for
$\theta_{13} > 0$ as well as the global bound, the latter being
dominated by other data sets.

The OPERA experiment~\cite{Acquafredda:2009zz} at the CERN to Gran Sasso
(CNGS) beam has an excellent electron neutrino selection efficiency, but the
beam setup is optimised for \numu--\nutau transition searches, being not
very competitive for electron neutrino appearance searches. OPERA
sensitivities to $\theta_{13}$ have been firstly computed
in~\cite{Komatsu:2002sz} and then in \cite{Huber:2004ug} where a sensitivity
of $\sin^2{2\thetaot} \geq 0.14$ (90\% C.L., for $\Delta m^2_{31}=10^{-3}$
eV$^2$ ) has been estimated in a 5 years neutrino run at the nominal CNGS
intensity of $4.5 \cdot 10^{19}$ pot/year (expected to happen after 2013).
Possible upgrades of CNGS have been studied in
\cite{Meregaglia:2006du,Baibussinov:2007ea}
where an off-axis liquid argon detector would detect neutrinos
at the first oscillation maximum.
For a discussion of these experimental possibilities see
\cite{Battiston:2009ux} and also section~\ref{sec:next}.

\bigskip {\bf Note added:} After the completion of this review the $\nu_e$
appearance data from MINOS corresponding to $7\times 10^{20}$~pot has been
made public~\cite{Adamson:2010uj}. 54 events are observed with an expected
background of $49.1\pm 7.0\pm 2.7$. The $1.5\sigma$ excess found initially
in~\cite{Collaboration:2009yc} has reduced now to about $0.7\sigma$ and data
are therefore in perfect agreement with background expectations for no
$\nu_e$ appearance. The results and figures in this section correspond to
the initial $3.14\times 10^{20}$~pot, while final results on $\theta_{13}$
in the global analysis from \cite{Schwetz:2008er} have been updated with the
$7\times 10^{20}$~pot data.

\subsection{$\theta_{13}$ in the global three--flavour fit}
\label{sec:global}

Let us now summarise the present situation obtained in the global fit of all
relevant oscillation data, as illustrated in fig.~\ref{fig:th13-bound}. In
the updated analysis\footnote{The results presented here are based on the
arXiv version 3 of~\cite{Schwetz:2008er}, updated with the $7\times
10^{20}$~pot MINOS $\nu_e$ appearance results from~\cite{Adamson:2010uj}.}
of~\cite{Schwetz:2008er} the following bounds at 90\% ($3\sigma$)~CL are
obtained (c.f.\ fig.~\ref{fig:th13-bound}, right):
\begin{equation}\label{eq:th13-2010}
  \sin^2\theta_{13} \le \left\lbrace \begin{array}{l@{\qquad}l}
      0.053~(0.078) & \text{solar+KamLAND} \\
      0.033~(0.058) & \text{CHOOZ+atm+K2K+MINOS} \\
      0.031~(0.047) & \text{global data}
    \end{array} \right.
\end{equation}
The ``hint'' for $\theta_{13} > 0$ coming from the different data sets can
be quantified by considering the $\Delta\chi^2$ for $\theta_{13} =0$:
\begin{equation}\label{eq:hint}
  \Delta\chi^2(\theta_{13} =0) = \left\lbrace 
  \begin{array}{l@{\quad}l@{\qquad}l}
      2.2 & (1.5\sigma) & \text{solar+KamLAND} \\
      0.8 & (0.9\sigma) & \text{CHOOZ+atm+K2K+MINOS} \\
      0.6 & (0.7\sigma) & \text{MINOS $\nu_e$ appearance}\\
      1.8 & (1.3\sigma) & \text{global data}
    \end{array} \right.
\end{equation}
%

\begin{table}
  \centering
  \begin{tabular}{lcc}
  \hline
    reference & best-fit and $1\sigma$ errors & significance \\
  \hline
    Fogli et al.~\cite{Fogli:2009ce} &
    $\sin^2\theta_{13} = 0.02\pm 0.01$ &
    $2\sigma$ \\
    Gonzalez-Garcia et al.~\cite{GonzalezGarcia:2010er} (GS98) &
    $\sin^2\theta_{13} = 0.0095^{+0.013}_{-0.007}$ &
    $1.3\sigma$ \\
    Gonzalez-Garcia et al.~\cite{GonzalezGarcia:2010er} (AGSS09) &
    $\sin^2\theta_{13} = 0.008^{+0.012}_{-0.007}$ &
    $1.1\sigma$ \\
    Schwetz et al.~\cite{Schwetz:2008er} (GS98) &
    $\sin^2\theta_{13} = 0.013^{+0.013}_{-0.010}$ &
    $1.5\sigma$\\
    Schwetz et al.~\cite{Schwetz:2008er} (AGSS09) &
    $\sin^2\theta_{13} = 0.010^{+0.013}_{-0.008}$ &
    $1.3\sigma$\\
  \hline
  \end{tabular}
  \caption{Comparison of the best-fit values for $\sin^2\theta_{13}$ and the
  significance of the hint for $\theta_{13} > 0$ from different global fits
  to neutrino oscillation data. The numbers from
  \cite{GonzalezGarcia:2010er} and \cite{Schwetz:2008er} include $7\times
  10^{20}$~pot $\nu_e$ appearance data from MINOS,
  whereas~\cite{Fogli:2009ce} is based on $3.14 \times 10^{20}$~pot. AGSS09
  and GS98 refer to low and high metalicity solar models,
  respectively~\cite{Serenelli:2009yc}. \label{tab:fit-comparison}}
\end{table}

In table~\ref{tab:fit-comparison} we compare the best-fit values for
$\sin^2\theta_{13}$ and the significance of the hint for $\theta_{13}
> 0$ from the global fits to neutrino oscillation data from three
different groups. All groups find a non-zero best-fit point in the
range $\sin^2\theta_{13} = 0.01 - 0.02$. Depending on the analysis as
well as variations in assumptions about the solar metalicity used as
input for the solar neutrino flux prediction a ``significance'' for a
non-zero $\theta_{13}$ between $1.1\sigma$ and $2\sigma$ is
found. Part of the differences can be attributed to differences in the
atmospheric neutrino analyses, as discussed in
section~\ref{sec:atm}. However, it is the opinion of the authors that
within the accuracy which one can attribute to global analyses of this
kind the results are in agreement. While it is premature to draw
strong conclusions from these results, upcoming experiments will answer
very soon the question whether $\theta_{13}$ is indeed in the range
indicated by present global analyses. This will be the topic of most
of the remaining part of this review.

Let us also mention that there are some limitations of the statistical
method applied to obtain these results. All three groups quoted in
table~\ref{tab:fit-comparison} use a $\Delta\chi^2$ method to evaluate
allowed regions as well as upper bounds on $\theta_{13}$ assuming
standard $\chi^2$-distributions. For example, the 90\%~CL bound is
obtained by the requirement $\Delta\chi^2 = 2.71$. The confidence
levels obtained by this method are only approximate close to the
physical boundary of a parameter, such as $\sin^2\theta_{13} \ge 0$ in
our case of interest. The confidence interval from such a
$\Delta\chi^2$ method should be interpreted as a two-sided confidence
interval, which lacks a well defined meaning close to the
boundary. Therefore, the results on $\theta_{13}$ quoted in this
section should be taken with some grain of salt, and in particular the
numbers given for various confidence levels (for the significance of a
hint for a non-zero $\theta_{13}$ as well as upper bounds on
$\theta_{13}$) have to be considered only as approximate, and should
always be understood in terms of the $\Delta\chi^2$ values. To convert
the $\Delta\chi^2$ values into a well defined probability would
require more elaborate statistical methods than used so far in the
literature.

\section{Description of upcoming experiments}
\label{sec:exp}

We move now to the discussion of the upcoming generation of
experiments searching for $\theta_{13}$, reviewing some general
principles of reactor experiments (sec.~\ref{sec:react}) and
accelerator neutrino beam experiments (sec.~\ref{sec:beams}). In both
cases we present the representatives of these classes of experiments:
Daya Bay~\cite{Guo:2007ug}, Double Chooz~\cite{Ardellier:2006mn},
RENO~\cite{RENO-cdr} for the reactor experiments, and
NO$\nu$A~\cite{Ambats:2004js} and T2K~\cite{Itow:2001ee} for the
beams.

\subsection{Reactor experiments}
\label{sec:react}

\begin{figure}
\centering \includegraphics[width=\textwidth]{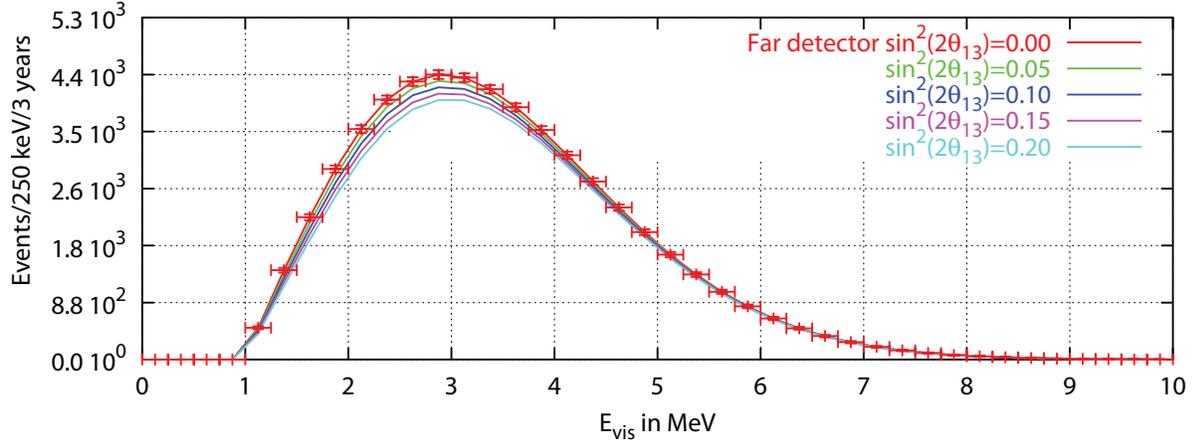}
  \caption{\label{fig:react-spect} Neutrino induced positron spectrum
  for various values of $\stheta$ simulated for the Double Chooz far
  detector~\cite{Ardellier:2004ui}.}
\end{figure}

Reactor experiments see a large signal of $\bar\nu_e$ events, and search for
a small deviation from the non-oscillation prediction due to
$\theta_{13}$-induced $\bar\nu_e$ disappearance, see
fig.~\ref{fig:react-spect}. This is a precision experiment, whose success
relies on statistical as well as systematical errors below the percent
level. 

The detection of nuclear reactor anti-neutrinos is a rather well known topic,
since neutrinos themselves have been discovered at a nuclear reactor in 1956 
\cite{Cowan:1992xc}.
Modern experiments are all inspired by the CHOOZ detector
\cite{Apollonio:1997xe}, and this technique is nicely described in
\cite{Bemporad:2001qy}. Here we will very briefly summarise the basic
principles.

The signal of a reactor anti-neutrino is provided by the reaction
\begin{equation}
  \nubare + p \rightarrow e^+ + n \,,
\end{equation}
a process with a threshold energy of 1.8~MeV where the positron provides a
prompt signal (boosted by the two 511 keV annihilation gamma rays) and the
neutron a delayed signal (in liquid scintillator this delay is about
170~$\mu$s). The coincidence of the two signals is powerful enough to make it
possible to run the experiment with a tolerable level of backgrounds.

Liquid scintillator can be loaded with 0.1\% natural gadolinium, an element
with high thermal neutron capture cross-section. In this way the neutron
capture time is reduced from $\sim 170 \, \mu$s to $\sim 27 \, \mu$s,
allowing for the reduction of the uncorrelated background. Furthermore, Gd
de-excitation after capture releases an 8 MeV $\gamma$ cascade (it would be
$\sim 2$ MeV in pure liquid scintillator), producing an integrated signal
well above the natural radioactivity.

The CHOOZ experiment concluded its data taking with a 2.8\% statistical and
a 2.7\% systematical error (c.f.\ eq.~\ref{eq:Chooz-result}). The goals of a
follow-up experiment are to improve CHOOZ sensitivity by a factor 5 at
least. This roughly reflects on a factor five improvement both in statistics
and in systematics.

Let's start on the easy part, the statistics. The CHOOZ detector was a 5~ton
detector exposed to two reactors of 8.6~MW thermal power at a distance of
1.05~km. The experiment integrated a total run of 8761.7~h, only 1543.1 of
which with the two reactors on and 3245.8~h with one of the two reactors on.
To gain a factor 25 in the number of neutrinos a detector twice as big
running 3 years with an improved efficiency is needed. The main limiting
factor in this direction is the stability of a gadolinium doped liquid
scintillation detector. Important progress has been made in this field in
recent years \cite{Hartmann} such that a running time of 5 or more years
seems feasible.

More generally the number of events can be estimated by
\begin{equation}\label{eq:react-events}
N \simeq 23\,000 
\left(\frac{L}{\rm km}\right)^{-2}
\left(\frac{T}{\rm yr}\right)
\left(\frac{P_{\rm th}}{\rm GW}\right)
\left(\frac{M}{100 \,\rm t}\right) \,,
\end{equation}
where $L$ is the baseline, $T$ is the running time of the experiment,
$P_{\rm th}$ is the thermal power of the nuclear power plant, $M$ is the
fiducial detector mass, and we have assumed 100\% efficiency.

More difficult is the reduction of systematics. This task is particularly
challenging because CHOOZ had the very rare opportunity of directly
measuring the backgrounds with the nuclear reactors off. For a complete
discussion and comparison of systematic errors at reactor experiments
see \cite{Mention:2007um}.

The first important action is to introduce a close detector as similar as
possible to the far detector \cite{Mikaelyan:1999pm,
Martemyanov:2002td, Minakata:2002jv, Huber:2003pm}, in order to measure the
neutrino interaction rate before oscillations. In this way uncertainties on
the neutrino rates (around 2\%) almost cancel out. This approach has some
intrinsic limitations. The neutrino rates can't be the same in the two
detectors (even in absence of oscillation) because of the different coverage
and of the different distance from the source. The calibration and live-time
of the detectors must be kept as similar as possible. In a configuration
where more than one reactors are present, the close detector doesn't measure
the identical flux of the far one. This is because the neutrino flux of a
reactor varies in time according to the core composition which differs from
reactor to reactor following their fuel refurbishment.

The backgrounds of reactor neutrino experiments are of two types:
uncorrelated signals from cosmic rays and natural radioactivity and
correlated signals from neutrons generated by cosmic muons. Uncorrelated
signals can be separately measured and normalised. Their random coincidences
can generate background events that can be kept at a negligible level with a
careful choice of low activity materials and design of the detector.

More problematic are neutrons induced by cosmic muons. A cosmic muon not
crossing the detector can induce spallation or be captured on the materials
outside the detector. These can produce neutrons that escape the vetoes and
produce correlated signals inside the detector that can mimic the
anti-neutrino signals. Among the cosmogenically produced isotopes there are
some long lived, as $^8$He and $^9$Li, with decay times of 119~ms and 174~ms
respectively~\cite{Hagner:2000xb}, that make hardware active vetoes
impractical. As an example, the close detector of RENO is expected to have a
25\% dead time for a 0.5 ms veto after any detected muon in the outer veto.
This is the main reason that forces detectors to run at shallow depth,
complicating very much the choice of the possible sites.

\begin{figure}
 \centering 
 \includegraphics[width=0.55\textwidth]{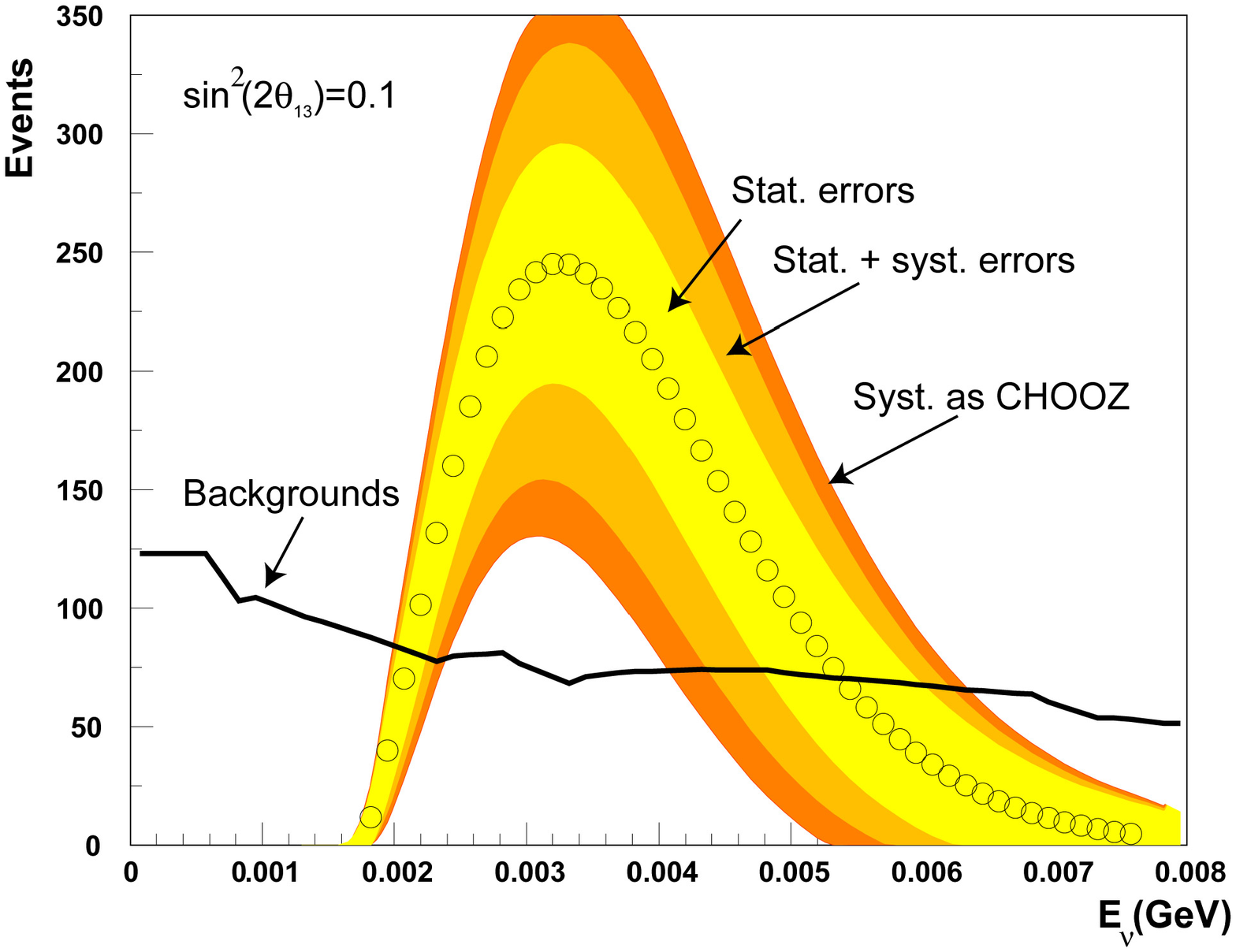}
 \caption{\label{fig:DC-signal} Number of disappeared events in Double Chooz
 for $\stheta = 0.1$, i.e., the difference between no-oscillation and
 oscillation spectra in the far detector, compare
 fig.~\ref{fig:react-spect}. Also shown are statistic and expected
 systematic errors, together with systematic errors as big as the former
 Chooz experiment. The signal is compared to the expected background rate.
 Based on data from~\protect\cite{Ardellier:2006mn}.}
\end{figure}

Fig.~\ref{fig:DC-signal} illustrates the size of the disappearance signal in
Double Chooz, compared to statistical and systematical errors as well as the
expected background.
Even if clearly inspired by the CHOOZ design, next generation reactor
neutrino detectors have introduced improvements in the design in order to
reduce the primary sources of systematics: knowledge of the fiducial volume
and backgrounds.

\begin{figure}
\centering
\includegraphics[width=105mm]{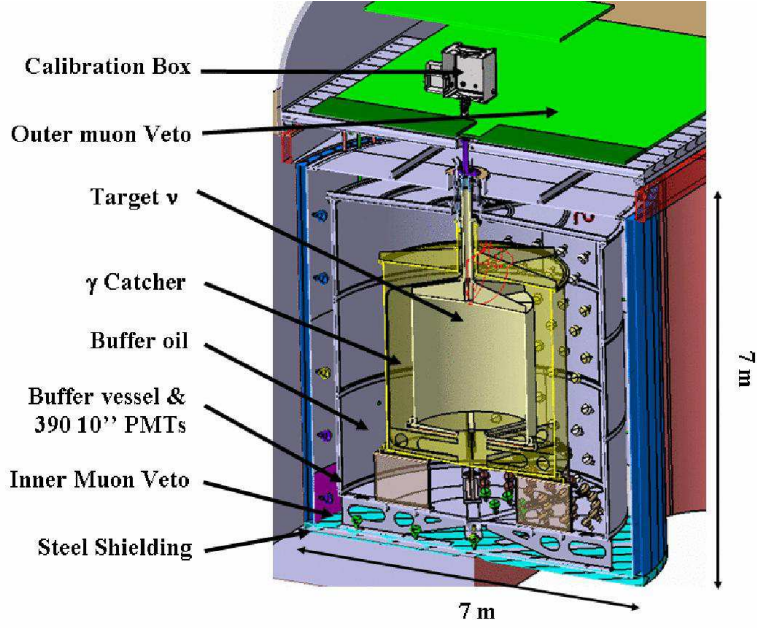}
\caption{\label{fig:detector}Sketch of the Double Chooz detector
design~\cite{Ardellier:2006mn}.}
\end{figure}

In the following, as an example, we will describe the detector of the Double
Chooz experiment, shown in fig.~\ref{fig:detector}. The inner detector is
made of 10.3 m$^3$ (8.3 metric tons, with a diameter of 2.3 m) of Gd-loaded
(0.1\%) liquid scintillator in an acrylic vessel. It is immersed in a gamma
catcher of 22 m$^3$ (3.4 m of diameter) of undoped scintillator, aimed to
detect the gammas emitted in both the neutron-capture process and positron
annihilation in the target. In this way gammas emitted from signal neutrino
events in the outer volume of the target are detected, providing a
well-defined target volume. A third shield of non-scintillating paraffin
oil, 5.5~m diameter, separates the active target and the gamma catcher from
the photo-multipliers, greatly reducing the intrinsic radioactivity of the
390 10-inch photo-multipliers,
the most radioactive component of the detector (this third shield was not
installed inside the CHOOZ detector). The outer detector volume is steel
walled, 6.6 m diameter, filled with scintillator and lined with 70 8-inch
photo-multipliers, 
equipping the Inner Veto, having the purpose of detecting and tracking muons
and fast neutrons. An Outer Veto is placed on top of the detector, made of
strips of plastic scintillator and wavelength-shifting fibres. Its task is
to tag muons interacting around the detector producing cosmogenic isotopes,
some of which produce correlated backgrounds in the inner detector, the most
dangerous source of backgrounds for the experiment.

\begin{table}
\begin{tabular}{lcccrcc}
\hline
Setup &  $P_{\mathrm{Th}}$ [GW] & $L$ [m] & $m_{\mathrm{Det}}$ [t] & Events/year & Backgrounds/day\\
\hline
 Daya Bay~\cite{Guo:2007ug} &  17.4  & 1700  &   80\hspace{1ex}  & $10 \hspace{1.7ex}\cdot 10^4$ & 0.4 \\
 Double Chooz~\cite{Ardellier:2006mn} &  \hspace{1ex}8.6  & 1050 &   \hspace{1ex}8.3 & $  1.5 \cdot 10^4$ & 3.6 \\
 RENO~\cite{RENO-cdr}    &  16.4  & 1400  &  15.4  & $3\hspace{1.7ex} \cdot 10^4$ & 2.6 \\
\hline
\end{tabular}
\caption{\label{tab:exp-summary} Summary of experimental key parameters of
upcoming reactor neutrino experiments. We give the thermal reactor power,
the approximate distance between reactors and far detector, and detector
mass, neutrino events per year, and background events per day, all for the
far detector. RENO backgrounds are the sum of correlated backgrounds as
computed in \protect \cite{RENO-cdr} and uncorrelated backgrounds as
estimated in \protect \cite{Mention:2007um}.}
\end{table}

\begin{figure}
  \includegraphics[width=\textwidth]{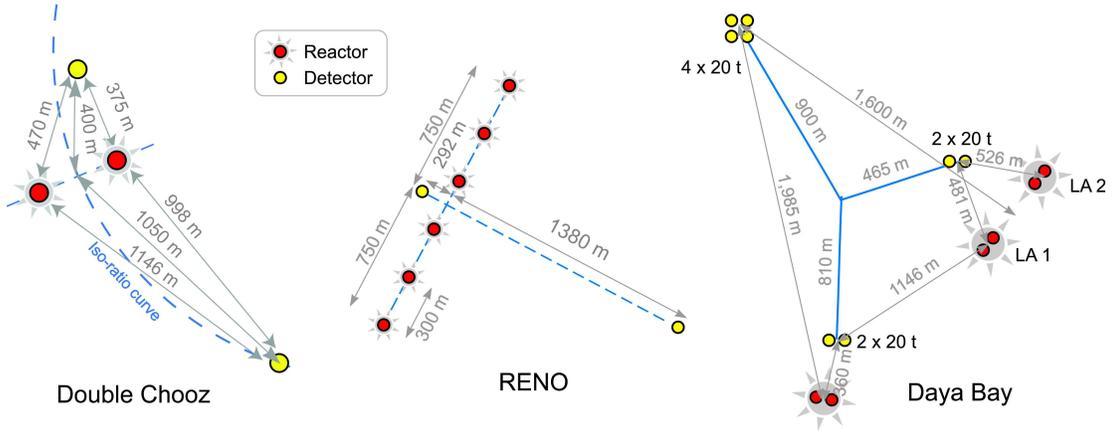}
  \caption{\label{fig:react-cores} Configuration of the experimental layout
  of Double Chooz, RENO, and Daya Bay. The dashed curve in the Double Chooz
  configuration is the far flux iso-ratio curve. Updated from \protect
  \cite{Mention:2007um}.}
\end{figure}

Below we describe briefly the three next generation reactor
experiments. Table~\ref{tab:exp-summary} summarises a few key
parameters and fig.~\ref{fig:react-cores} illustrates the
reactor/detector configurations.

\subsubsection{Double Chooz}

The Double Chooz experiment \cite{Ardellier:2006mn} is being installed near
the Chooz two-core (4.27+4.27 GW) nuclear power plant. The far detector,
described in the previous section, is at 1.05 km from the two reactor cores,
in the same site as the original CHOOZ experiment, at a depth of about 300
m.w.e.\ (meters of water equivalent). The close detector is designed to be
identical to the far detector, it is placed 400 m from the two reactors, at
a depth of 115 m.w.e., see fig.~\ref{fig:react-cores}. Let us mention that for
$\dma \simeq 2.5\times 10^{-3}\,\eVq$ the far detector distance of 1~km is
somewhat too short, shifting the oscillation signal to the lower part of the
spectrum, whereas the near detector distance of 400~m is somewhat too far,
with some effect of oscillations already present, see
e.g.~\cite{Huber:2003pm} for far and near detector baseline optimisation
studies. 

The number of neutrinos detected in the far detector assuming 3 years
running time will be $\simeq 45000$ compared to 2700 in the Chooz
experiment, reducing the statistical error from 2.8\% to 0.47\%. The goal
about systematic errors is to reach a level of 0.6\%. Without the near
detector the systematic error would be about 2.5\%.

\subsubsection{Daya Bay}

The Daya Bay experiment \cite{Guo:2007ug} will receive neutrinos from three
nuclear plants, each consisting of two cores: Daya Bay, Ling Ao I and Ling
Ao II (scheduled to be commissioned by the end of 2010) located in the south
of China, 55 km to the northeast of Hong Kong. The thermal power of each
core is 2.9 GW, hence the existing total thermal power is 11.6~GW, and will
be 17.4~GW after 2010.

The basic experimental layout of Daya Bay consists of three underground
experimental halls, one far and two close, linked by horizontal tunnels
under construction. The geometry of the reactor cores, the two near detector
stations and the far detector station is illustrated in
fig.~\ref{fig:react-cores}. It is evident from this configuration that the
flux measured by the near detectors will not be identical to the one
seen by the far detector. Each close detector hall, at a depth of about
100 m.w.e., will host two 20 ton gadolinium doped (0.1\%) liquid
scintillator detectors, while the far hall, 350 m.w.e., will host four such
detectors. The detectors are designed to be movable so that the close
detectors and the far ones can be swapped. This swapping is not necessary to
reach the designed systematics (0.38\%), but could be performed to cross
check the sensitivity and possibly further reduce the systematic errors
(down to 0.18\%).

\begin{figure}
\begin{center}
\includegraphics[width=0.5\textwidth]{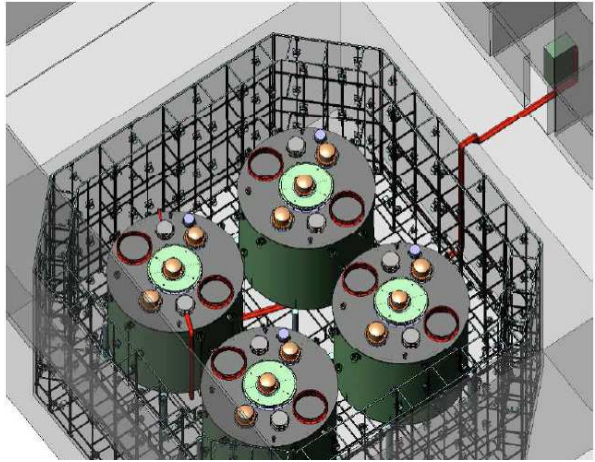}
\caption{A drawing of the Daya Bay far detector hall showing the four detectors in
a water pool instrumented with PMTs to detect cosmic muons.
From~\cite{Guo:2007ug}.
\label{fig:muon}}
\end{center}
\end{figure}

The design of the detectors is very similar to the Double Chooz one, with
the notable difference that the three stations will be submerged in a water
pool of 2.5~m depth to shield the detectors from ambient radiation and
spallation neutrons, see fig.~\ref{fig:muon}. Above the pool a muon tracking
detector made of 4 layers of resistive-plate chambers (RPCs) will be
installed.

\subsubsection{Reno}

The RENO~\cite{RENO-cdr} experiment is located on the site of the Yonggwang
nuclear power plant in the southwestern part of Korea. The plant consists of
six reactors lined up in roughly equal distances and spans about 1.3~km.
With a total thermal power of 16.4~GW it is at present the second largest
nuclear reactor plant in the world.

RENO will use two identical detectors, a close detector at about 290~m from
the reactor array (at a depth of about 110 m.w.e.) and a far detector at
1380~m (at a depth of about 450 m.w.e.), see fig.~\ref{fig:react-cores}.
The design of the RENO detectors is very similar to Double Chooz, the active
target will be 15.5~ton of Gd loaded (0.1\%) liquid scintillator. The outer
veto system is a layer of water, 1.5~m thick, contained in a 30 cm thick
concrete vessel. PMTs are mounted on the inner surface of the veto container
for detecting \ceren light from cosmic muons.
Goal of the experiment is to reach a $<0.5\%$ systematic error.

\subsection{Accelerator experiments}
\label{sec:beams}

Accelerator experiments look for the appearance of the $\nu_e$
flavour in an almost pure \numu beam, due to oscillations. 
The background rate has an intrinsic component given by the \nue
contamination in the neutrino beam, ranging from 0.5\% to 1\%.
Detector backgrounds can be generated by NC events where a $\pi^\circ$ produces
a signal misidentified as an electron, or \numu CC events where the muon
is misidentified as an electron. 

Several experiments already tried in the past to detect \numunue transitions
at accelerators with short baselines. The most stringent result comes from
the NOMAD experiment that reached a sensitivity on the probability of these
transitions $P(\numunue) \leq 0.7 \cdot 10^{-3}$ (90\% CL)
\cite{Astier:2003gs}.  The challenge of the next generation of long baseline
experiments is to obtain similar sensitivities with detectors $10^4$ more
massive than the NOMAD detector \cite{Altegoer:1997gv}, a 3~ton, low density
(0.1 g/cm$^3$), very sophisticated spectrometer.
The pioneering \numunue long-baseline appearance searches have been
performed by the K2K~\cite{Ahn:2004te} and
MINOS~\cite{Collaboration:2009yc} experiments, see
section~\ref{sec:minos}.

A rough estimate for the sensitivity can be obtained by comparing the signal
to the statistical and systematical uncertainty of the background:
\begin{equation}\label{eq:sig_ov_back}
\frac{S}{\sqrt{B + \sigma_{\rm bg}^2 B^2}} \,,
\end{equation}
where $S~(B)$ is the number of signal (background) events, and $\sigma_{\rm
bg}$ is the uncertainty in the background.

\begin{figure}
 \centering 
 \includegraphics[width=0.6\textwidth]{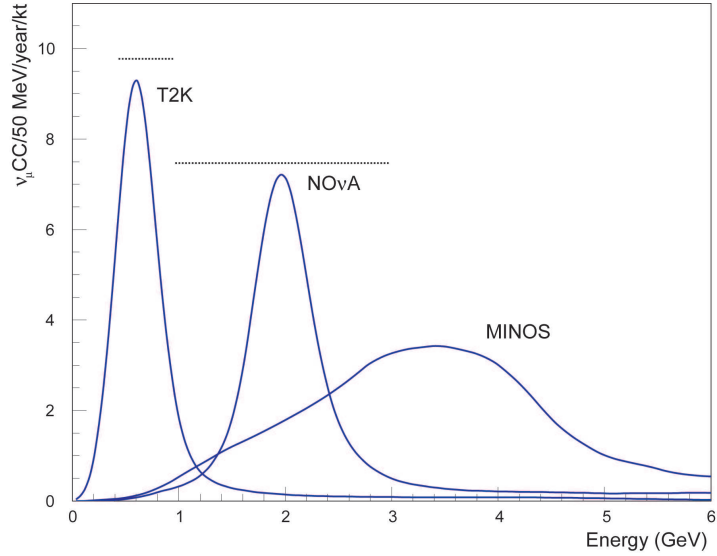}
 \caption{\label{fig:numucc-rates} Muon neutrino interaction rates in MINOS,
 NO$\nu$A, and T2K assuming no oscillations, normalised to one year of
 operation at the power of 410, 700, and 750~KW, respectively. The
 horizontal dotted lines represent the neutrino energy region where the
 \numunue oscillation probability is bigger than 50\% of the maximum
 ($\delta=0$ and normal hierarchy). The MINOS probability is very
 similar to NO$\nu$A. }
\end{figure}

It is almost impossible to derive a general rule for the number of events in
the far detector, since it depends very much on the geometry of the target,
the optics of the neutrino beam line, and the off-axis angle. The number of
signal events $S$ is proportional to the detector mass times the power of
the primary proton beam. Surprisingly enough $S$ does not depend very much
on the primary proton energy and on the baseline $L$ assuming the far
detector at the first oscillation maximum. This implies a constant value of
$L/E$ with $E$ the neutrino energy. Since the neutrino cross section
generally increases as $E^r$ with $r \simeq 1 - 2$, the $1/L^2$ scaling of
the beam flux is largely compensated. A comparison of the neutrino
interaction rates in T2K, NO$\nu$A and MINOS is reported in
fig.~\ref{fig:numucc-rates}.

The SuperKamiokande water \ceren detector used in T2K already demonstrated
to be capable of keeping detector backgrounds to a level smaller than the
intrinsic \nue contamination of conventional neutrino beams. Also the total
active scintillator technology (TASD) chosen for NO$\nu$A allows to keep
beam and detector backgrounds at comparable level. As an example of signal
and background events in an accelerator experiment we reproduce signal and
background events in T2K as shown in \cite{Nakadaira}.

\begin{figure}
 \centering \includegraphics[width=0.6\textwidth]{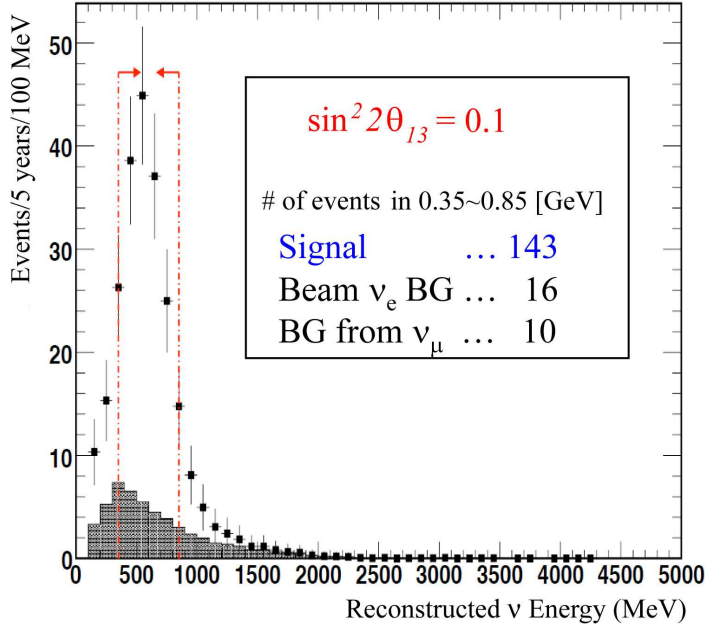}
 \caption{\label{fig:T2K-events} Signal and backgrounds events for T2K,
 computed for $\stheta=0.1$, $\delta=0$, $\Delta m^2_{31}=3\cdot
 10^{-3}$~eV$^2$ and normal hierarchy. From~\protect\cite{Nakadaira}.}
\end{figure}

However, the water \ceren and TASD techniques allow a reasonable event
energy reconstruction only for quasi-elastic (QE) neutrino interactions. QE
interactions produce just one charged particle above the \ceren threshold,
resulting in just one \ceren ring in the detector. The two-body kinematics
allows a precise reconstruction of the incident neutrino energy from the
measured momentum of the outgoing lepton and the known direction of the
incoming neutrino. E.g., for a QE \numu event one has
\begin{equation}
E_\nu^{\rm rec}\simeq \frac{1}{2}\frac{(M_p^2-m^2_\mu)+2E_\mu M_n-M_n^2}
{-E_\mu+M_n+p_\mu \cos{\theta_\mu}}
\end{equation}
where $M_p$, $M_n$, $m_\mu$, $E_\mu$, $p_\mu$, $\cos{\theta_\mu}$ are the
proton, neutron, muon masses, muon energy, momentum and angle with respect
to the incoming neutrino direction. Single ring events produced by non-QE
interactions could bias the energy reconstruction because in their case the
two body kinematics does not hold. For this reason it is important to
precisely know the non-QE/QE ratio in the neutrino interactions.
 
Other detector technologies, such as liquid argon TPCs, could suppress
detector backgrounds to a negligible level and efficiently reconstruct
neutrino interaction events of any multiplicity, and are very promising for
future neutrino oscillation experiments.

A beam configuration used to optimise long baseline experiments with the
goal of measuring \thetaot is the off-axis configuration, a concept that was
firstly proposed in \cite{Mann:1993gi}. According to the two-body
$\pi$-decay kinematics, all the pions above a given momentum produce
neutrinos of similar energy at a given angle $\theta \ne 0$ with respect to
the direction of the parent pion (contrary to the $\theta=0$ case where the
neutrino energy is proportional to the pion momentum).

An off-axis configuration offers several advantages for a long baseline
experiment optimised to detect sub-leading \numunue oscillations. First, the
off-axis neutrino flux at the desired energy is higher than in an on-axis
configuration. Second, the neutrino flux at higher energies than the
oscillation maximum is greatly reduced, with a consequent reduction of
backgrounds due to $\pi^\circ$ generated by neutral current events. Both of
these effects are clearly visible in comparing the MINOS (on-axis) and
NO$\nu$A (off-axis) spectra in fig.~\ref{fig:numucc-rates}. Third, the
intrinsic background due to beam \nue\ is less around the oscillation
maximum (about $0.4\%$ in T2K) since \nue\ are mostly generated by three
body meson decays that have different kinematics than the two body pion
decays. These advantages overcompensate the smaller total neutrino flux
(even when multiplied by the oscillation probability) of the off-axis
configuration.

In an appearance experiment the influence of systematic errors is very much
reduced if compared to the reactor experiments. Nevertheless a close
detector is necessary in order to precisely measure the flux of \numu, that
are going to oscillate into \nue in the far detector, to measure interaction
cross sections of signal and background processes, and to measure the
backgrounds for the signal \nue. As extensively discussed in
\cite{Huber:2007em} the task of the close detector is not just a background
subtraction, since for instance, factors as the QE/non-QE ratio can spoil
the experimental sensitivity. As it will be described in the following the T2K
experiment renounced to have a replica of the far detector in the close
position, and designed a very sophisticated close detector capable of
precisely measuring the exclusive neutrino cross sections that play a role
in the signal and backgrounds in the far detector as well as the flux of
\numu.
NO$\nu$A instead is planning to have just a replica of the far detector in
the close station, but it shouldn't be missed the fact that in the same beam
line (on-axis) will soon operate experiments specialised in measuring
neutrino cross sections, like MINER$\nu$A \cite{Drakoulakos:2004gn}.

\subsubsection{T2K}

\begin{figure}
\centering 
   \includegraphics[width=0.6\textwidth]{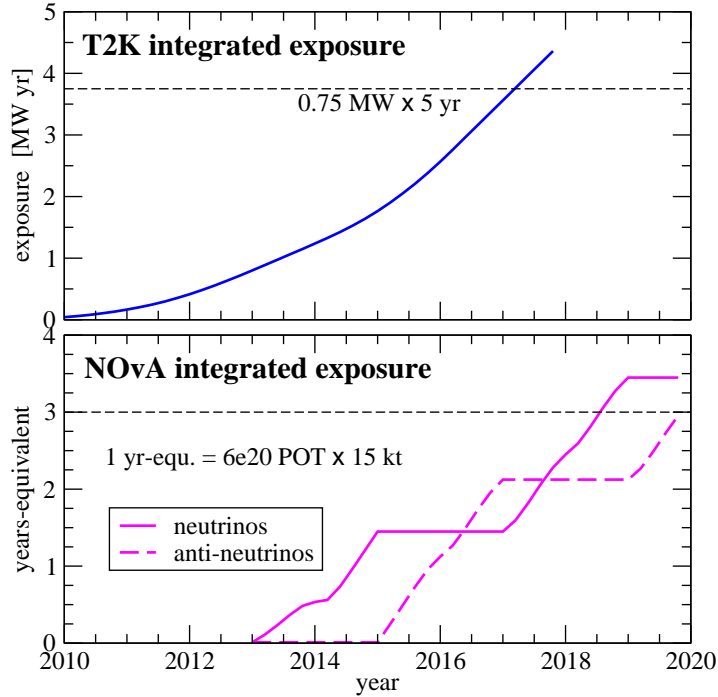}
   \caption{Integrated exposures for the T2K (top) and NO$\nu$A
   (bottom) experiments. The T2K beam power plan is based
   on~\cite{Nakadaira}. The NO$\nu$A exposure has been normalised to
   nominal ``NO$\nu$A years'', which corresponds to $6\times 10^{20}$
   protons-on-target (pot) and a 15~kt detector.  The shown curves
   take into account the plan for the available detector mass as a
   function of time, as well as the available pot~\cite{messier}.}
 \label{fig:beam-exposure}
\end{figure}

The T2K (Tokai--to--Kamioka) experiment~\cite{Itow:2001ee} will use a high
intensity off-axis ($2.5^\circ$) neutrino beam, with a peak energy of 700
MeV, generated by a 30 GeV proton beam at J-PARC (Japan Proton Accelerator
Research Complex) fired to the SuperKamiokande detector, located 295 km
from the proton beam target. 
The design intensity of the J-PARC proton beam is $3.3\cdot10^{14}$
protons/pulse with a repetition rate of 3.3 s, corresponding to a beam power
of about 750 kW. 
The experiment is expected to collect data for about $10^7$ s/year, 
performances are normalised to a 5 year run at the nominal power.
Fig.~\ref{fig:beam-exposure} shows the integrated exposure as a function of
time using the expected beam power shown in~\cite{Nakadaira}. According to
this plan, the nominal exposure will be reached around 2017.

\begin{figure}
\centering 
 \includegraphics[width=0.6\textwidth]{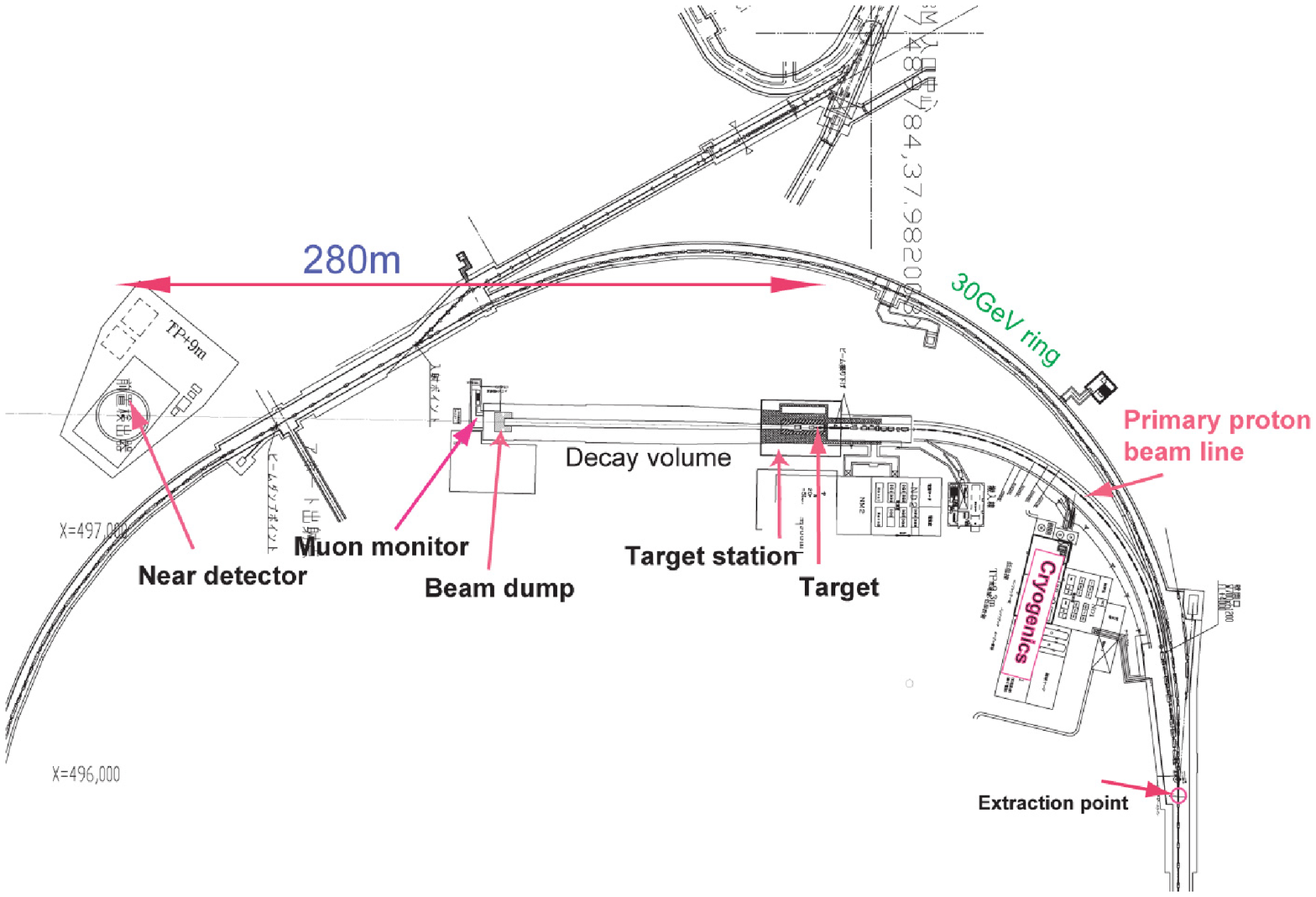}
 \includegraphics[width=0.38\textwidth]{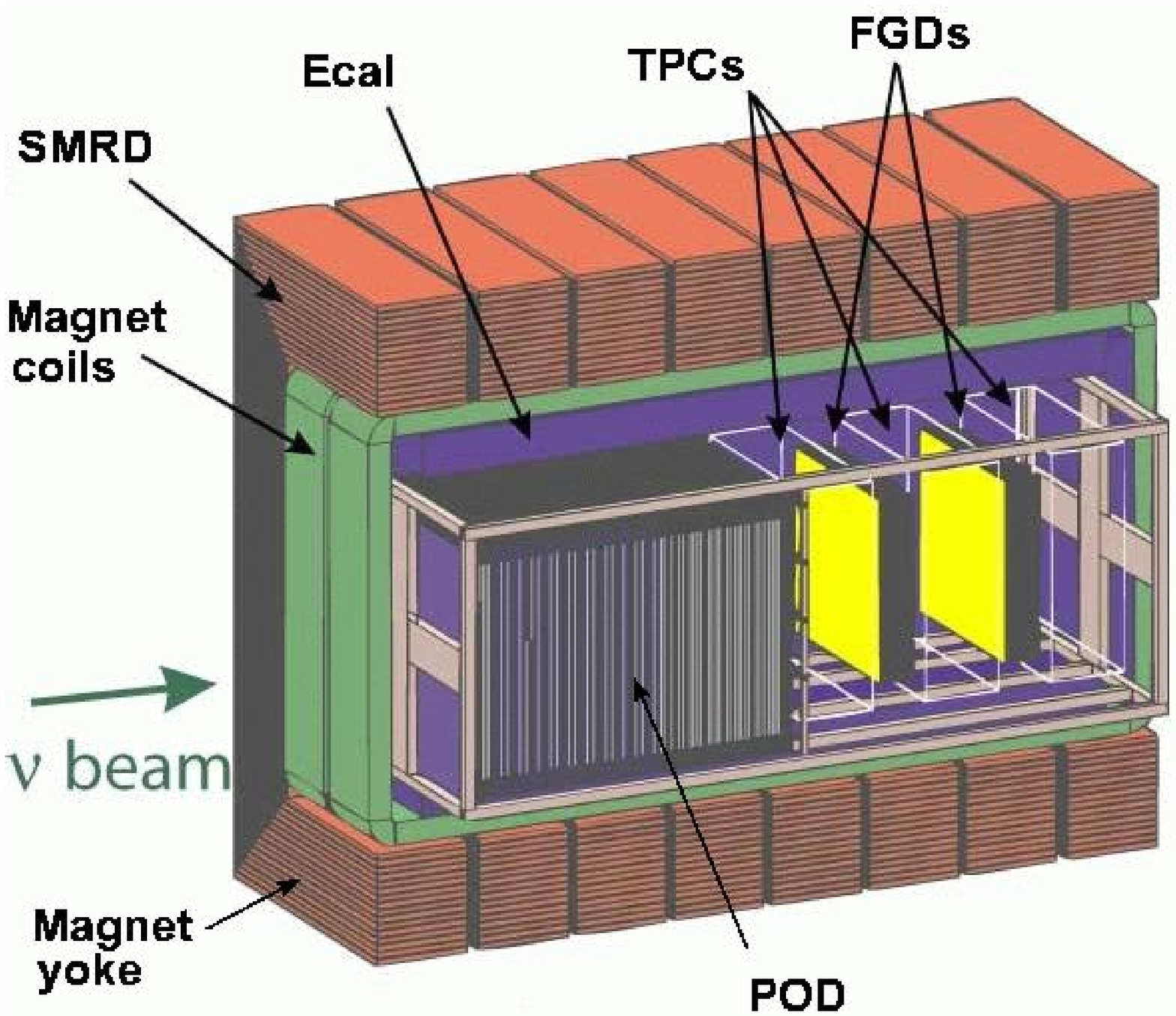}
 \caption{Left: the layout of the T2K beam line, showing the location of
 primary proton beam line, target station, decay volume, beam dump, muon
 monitors and near neutrino detectors,\protect \cite{Sekiguchi:2008zza}.
 Right: sketch of the T2K ND280 off-axis near detector.  \label{fig:t2k-beamline}}
\end{figure}

The schematic view of the T2K neutrino beam line is shown in
fig.~\ref{fig:t2k-beamline} (left). The primary proton beam interacts in a
helium-cooled graphite target placed inside the first horn. The beam optics
includes three horns running at 320 kA with a maximum magnetic field of 2.1
T. The decay volume is 94 m long after the target region with a variable
cross section starting at 2.2 m (W) at 2.8 m (H) and increasing to 3.0~m at
4.6~m at the far end, filled with inert helium gas held at 1 atm. A beam
dump, constructed of cooled graphite and copper blocks is placed at the end
of the decay volume.

A sophisticated near detector complex has been built at a distance of 280 m
from the target.  This complex has two detectors: one on-axis (INGRID) and
the other off-axis (ND280). This off-axis detector
(fig.~\ref{fig:t2k-beamline}, right) is a spectrometer built inside the
magnet of the former experiments UA1 and NOMAD, operating with a magnetic
field of 0.2~T. It includes a Pi-Zero detector (POD), a tracking detector
made by three time projection chambers (TPCs) and two fine grained
scintillator detectors (FGDs) acting as an active target, a $4\pi$
electromagnetic calorimeter (Ecal), and a side muon range detector~(SMRD).  
To reduce systematic errors both the POD and the FGDs incorporate water
targets (the same material of the far detector).
Neutrino rates in the close detector will be about 160000 \numu (3200 \nue)
interactions/ton/yr at the nominal beam intensity of 0.75~MW$\cdot 10^7$~s.

ND280 is expected to calibrate the absolute energy scale of the neutrino
spectrum with 2\% precision, measure the non-QE/QE ratio at the 5-10\% and
monitor the neutrino flux with better than 5\% accuracy. The momentum
resolution of muons from the charged current quasi-elastic
interactions~(CCQE) should be better than 10\%.  The $\nu_e$ fraction should
be measured with an uncertainty better than 10\%. A measurement of the
neutrino beam direction, with a precision better than 1 mrad, is required
from the on-axis detector.  

The close detector location, 280 m from the target, is too close to the
decay tunnel to have a neutrino flux identical to the far detector flux.
Indeed differences as big as 50\% are expected between the two fluxes,
reducing the capability of the close detector of reducing systematic errors
independently from any Monte Carlo simulation. A more distant close detector
station could attenuate the near-far neutrino flux differences, but for
the moment is not foreseen by the experiment.

A fundamental tool to better control the neutrino flux is the ongoing
measurement of the hadro-production in a T2K target replica by 30 GeV protons
by the NA61 experiment at CERN \cite{Posiadala:2009zf}.  Following the
succesful example of the measurements done by the HARP experiment
\cite{Catanesi:2007ig} for the K2K \cite{Catanesi:2005rc} and for the
MiniBooNE \cite{Catanesi:2007gt} targets, NA61 data should greatly improve
the accuratness of the neutrino beam line simulation.

Signal events are detected by the SuperKamiokande detector with an
efficiency of 45\% and with a detector background contamination smaller than
the intrinsic beam \nue contamination (roughly in the 2:3 ratio, 
see also fig.~\ref{fig:T2K-events}).

\subsubsection{NO$\nu$A} 

The NO$\nu$A experiment~\cite{Ambats:2004js} will run at an upgraded NuMI
neutrino beam expected to deliver $6.5 {\cdot} 10^{20}$ pot/year,
corresponding to a beam power of 700 kW, generating a neutrino beam with an
average energy $E_{\nu} \sim 2 $ GeV and a $\nu_e$ contamination less than
$0.5 \%$.

\begin{figure}
\centering
 \includegraphics[width=0.6\textwidth]{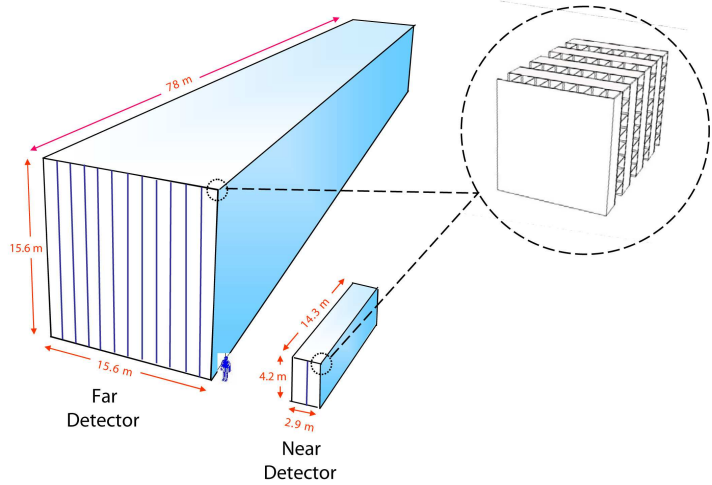}
 \caption{Sketch of the NO$\nu$A detectors. The insert figure shows
  the plane arrangement of each detector.
  From \protect \cite{Nova-web}.}
\label{fig:nova-detectors}
\end{figure}

The far detector, placed at baseline of 810~km, 14 mrad (0.8$^\circ$)
off-axis, will be a ``totally active'' tracking liquid scintillator,
constructed from liquid scintillator contained inside extruded PVC cells.
Each cell is 3.9 cm wide by 6.0 cm deep and is 15.5 meters long, they will
be arranged in an alternating plane structure composed of vertical and
horizontal cells. Scintillation light will be guided to APD photo-detectors
using wavelength shifting fiber.
NO$\nu$A will be located near Ash River and will have a total mass of 15
kilotons (385000 cells) and be 15.7 meters wide, 15.7 meters tall, 78 meters
long (the only non-cylindrical detector in this generation of experiments).
It will run on surface, with a modest overburden of 3~m of concrete. It is
expected to detect \nue signals with an efficiency of 26\% (including
acceptance) and to reduce the detector background rate to a level comparable
to the rate from the intrinsic beam \nue contamination. 
The close detector will be a 215 ton replica of the far detector, placed 14
mrad off the NuMI beam axis at a distance of 1 km from the target. A sketch
of the NO$\nu$A detectors is shown in fig.~\ref{fig:nova-detectors}.

NO$\nu$A plans to alternate between neutrino mode and antineutrino mode: the
focus of the experiment is to provide data on the neutrino mass hierarchy,
where NO$\nu$A has a clear advantage with respect to T2K thanks to the
longer baseline, see e.g.~\cite{Huber:2009cw} for a recent sensitivity
study, and \cite{Huber:2002rs, Minakata:2003ca, Mena:2006uw} for discussions
of the T2K/NO$\nu$A complementarity/synergy. A possible
scenario~\cite{messier} on the time evolution of neutrino and antineutrino
exposure is shown in fig.~\ref{fig:beam-exposure}, which takes into account
the build-up of the detector mass as well as the available beam power.
As a second phase, the NuMI beam intensity could be increased to 1.2 MW
(``SNuMI'') or to 2.3 MW (``Project X'') in case the new proton driver of 8
GeV/c and 2 MW will be built at FNAL.

%
\section{Phenomenology and sensitivity estimates for upcoming experiments}
\label{sec:pheno}

\subsection{Complementarity of reactor and superbeam experiments}
\label{sec:complementary}

A natural classification of the 5 upcoming experiments is the
distinction between reactor and accelerator (superbeam)
experiments. Apart from the vastly different experimental
configurations and challenges (see section~\ref{sec:exp}) there is
also an important difference from the point of view of oscillation
physics. While reactor experiments search for the disappearance of
electron antineutrinos, accelerator experiments look for the
appearance of electron (anti)neutrinos in a beam initially composed
mainly of muon (anti)neutrinos.

Let us now discuss the relevant oscillation probabilities. We denote with
$L,E_\nu,V$ the distance from the neutrino source and the detector, the
neutrino energy, and the matter potential, respectively, with $V = \sqrt{2}
G_F N_e$ where $N_e$ is the electron density along the neutrino path. 
Furthermore we introduce the abbreviations
\begin{equation}
\Delta \equiv \frac{\Delta m^2_{31} L}{4 E} \,,\quad
A \equiv \frac{2EV}{\Delta m^2_{31}} \,,\quad
\alpha \equiv \frac{\dms}{\dma} \,.
\end{equation}
Note that the signs of all three of these quantities depend on the neutrino
mass hierarchy encoded in the sign of $\dma$. 

In the case of reactor experiments with $L \lesssim 2$~km the matter effect is
negligible and the survival probability is given by
\begin{equation}\label{eq:prob-react}
P_{ee} \approx 1 - \stheta \sin^2 \Delta 
- \alpha^2 \Delta^2 \sin^22\theta_{12}  \,.
\end{equation}
As mentioned in the discussion of the CHOOZ experiment in
section~\ref{sec:chooz}, the last term due to $\dms$ is suppressed by the
small solar mass squared difference and can be neglected as long as $\stheta
\gg \alpha^2 \simeq 10^{-3}$, and in this case $P_{ee}$ is independent
of the solar parameters $\theta_{12}$ and $\dms$. Furthermore, the survival
probability eq.~\ref{eq:prob-react} does neither depend on $\theta_{23}$, on
the CP phase $\delta$, nor on the sign of $\dma$. Hence reactor experiments
provide a ``clean'' measurement of $\theta_{13}$, free of correlations with
other parameters, apart from $|\dma|$, which, however, is relatively
precisely known~\cite{Minakata:2002jv, Huber:2003pm}.

Let us now move to long-baseline appearance experiments. For typical
neutrino energies in the GeV range and baselines larger than 100~km
one cannot neglect the matter effect. However, for baselines smaller
than a few 1000~km it is a good approximation to assume a constant
matter density, given by the average density along the neutrino
path. In that case a rather useful expression for the
$P_{\nu_\mu\to\nu_e} \equiv P_{\mu e}$ appearance probability can be
obtained by considering terms up to second order in the small
quantities $s_{13}$ and $\alpha$ \cite{Cervera:2000kp, Freund:2001pn,
Akhmedov:2004ny}:
\begin{eqnarray}
  P_{\mu e} 
  &\approx \stheta \, s_{23}^2 \frac{\sin^2 (A-1)\Delta}{(A-1)^2} \nonumber\\
  &+ \alpha \, \sin 2\theta_{13} \, \sin 2\theta_{12} \, 
     \sin2\theta_{23} \cos(\Delta + \delta) \, 
     \frac{\sin A\Delta}{A} \, \frac{\sin (A-1)\Delta}{A-1}  \nonumber\\
  &+ \alpha^2 \, \sin^2 2\theta_{12} \, c_{23}^2 \frac{\sin^2 A\Delta}{A^2} \,. 
  \label{eq:prob-app}
\end{eqnarray}
This equation holds for neutrinos; for anti-neutrinos change $\delta
\to -\delta$ and $V \to -V$. The term in the first line of
eq.~\ref{eq:prob-app} is similar to a two--flavour oscillation
probability, apart from the $s_{23}^2$ factor which controls the
fraction of the $\nu_\mu$ flavour participating in the $\nu_e-\nu_\mu$
oscillations. This term dominates for large $\theta_{13}$. The term in
the second line of eq.~\ref{eq:prob-app} corresponds to an
interference term between oscillations with $\dma$ and $\dms$. It
depends on the CP phase $\delta$ and is responsible for CP
violation. The term in the last line is independent of $\theta_{13}$
and describes oscillations with $\dms$. It can be neglected for
$\stheta \gtrsim 0.01$.

\begin{figure}
\begin{center}
\includegraphics[width=\textwidth]{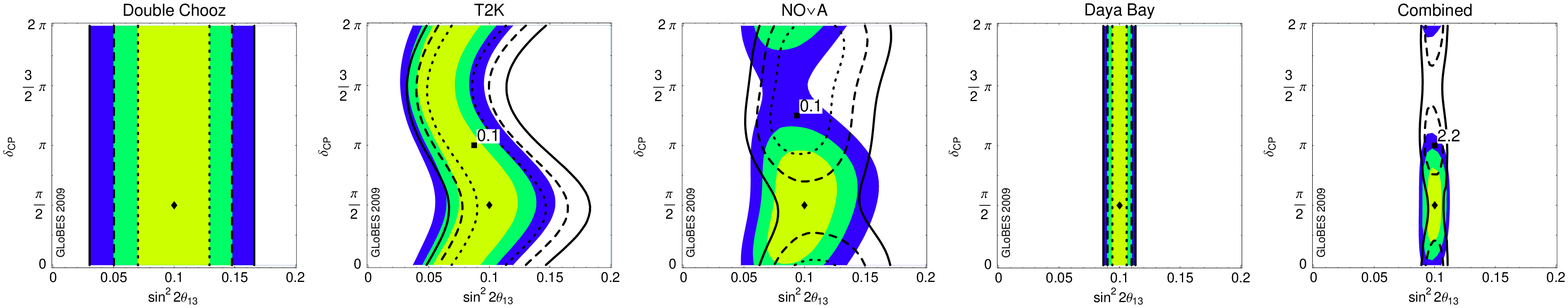} \\
\includegraphics[width=\textwidth]{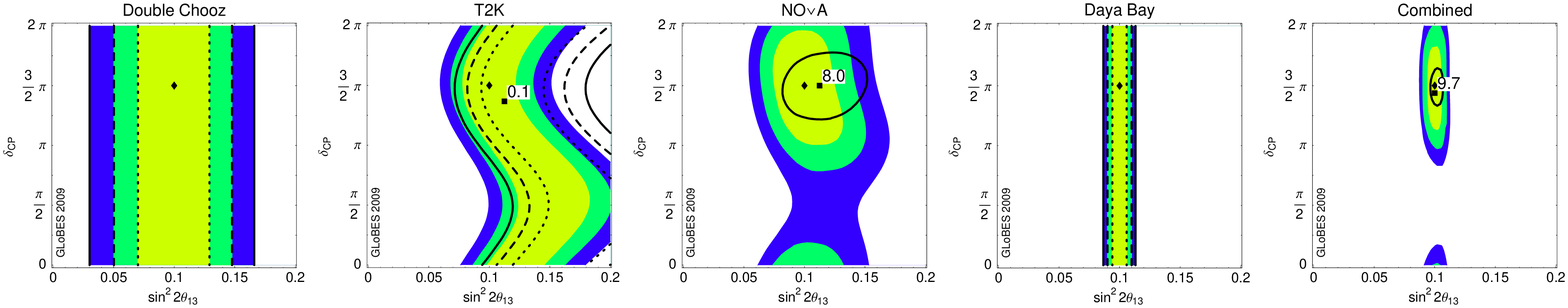}
\end{center}
  \caption{\label{fig:deltatheta} Exemplary fit results for
  Double Chooz, T2K, NO$\nu$A, Daya Bay, and the combination. Shown are
  fits in the $\theta_{13}$-$\delta$ plane assuming $\stheta=0.1$ and
  $\delta=\pi/2$ (upper row) and $\delta=3 \pi/2$ (lower row). A
  normal simulated hierarchy is assumed. The contours refer to
  $1\sigma$, $2 \sigma$, and $3 \sigma$ (2~dof). The fit contours for
  the right fit hierarchy are shaded (coloured), the ones for the wrong
  fit hierarchy are shown as curves. The best-fit values are marked by
  diamonds and boxes for the right and wrong hierarchy, respectively,
  where the minimum $\chi^2$ for the wrong hierarchy is explicitly
  shown. Reprinted from Ref.~\cite{Huber:2009cw}, Copyright (2009),
  with permission from JHEP.}
\end{figure}

Obviously, the parameter dependence of the appearance probability is
much more complicated than the one for disappearance in reactor
experiments, as it depends on all 6 oscillation parameters. If
information on $\theta_{13}$ is to be extracted from an appearance
measurement the correlations with the parameters $\delta$ and sgn($\dma$)
are especially important~\cite{Kajita:2001sb}.
The difference between reactor and super beam experiments is illustrated in 
fig.~\ref{fig:deltatheta}, taken from~\cite{Huber:2009cw}. It shows
how typical fits in the
$\theta_{13}$-$\delta$ plane would look like if $\theta_{13}$ was large
($\stheta=0.1$) and $\delta$ was close to maximal CP violation
$\delta=\pi/2$ (upper rows) and $\delta=3 \pi/2$ (lower rows). Here 
NH has been assumed to generate the ``data''. Using this ``true'' hierarchy
in the fit, the coloured regions are obtained. When the data is fitted with
the ``wrong'' hierarchy, i.e., with IH in this case, the regions delimited
by the curves are obtained.

The figures show the characteristics of the different classes of
experiments: The reactor experiments do not depend on $\delta$, and the
wrong fit hierarchy coincides with the right hierarchy. For T2K, which is
simulated with neutrino running only, there is some dependence on $\delta$,
but the correlation between $\delta$ and $\theta_{13}$ cannot be resolved.
We observe the typical
S-shape of the allowed regions, emerging from the $\cos(\Delta + \delta)$
term in eq.~\ref{eq:prob-app}. The wrong hierarchy contours are slightly
shifted, but the minimum $\chi^2$ is close to zero.  NO$\nu$A, on the other
hand, has both neutrino and anti-neutrino running in the simulation, which
means that the correlation can, at least in principle, be resolved. The
wrong hierarchy can in some cases be excluded because of larger matter
effects. In the combination of the experiments, the combination between Daya
Bay and the beams allows for a substantial reduction of the allowed
parameter space due to almost orthogonal measurements~\cite{Minakata:2002jv,
Huber:2003pm, McConnel:2004bd}. In the most optimistic cases, the mass
hierarchy can be determined at $3 \sigma$ confidence level, and maximal CP
violation can be established at relatively modest confidence as well.
However, note that these optimistic cases represent only a very small
fraction of the parameter space, see~\cite{Huber:2009cw} for a discussion of
the possibilities to measure CP violation and the mass hierarchy with these
experiments.

\begin{figure}
\centering \includegraphics[width=0.6\textwidth]{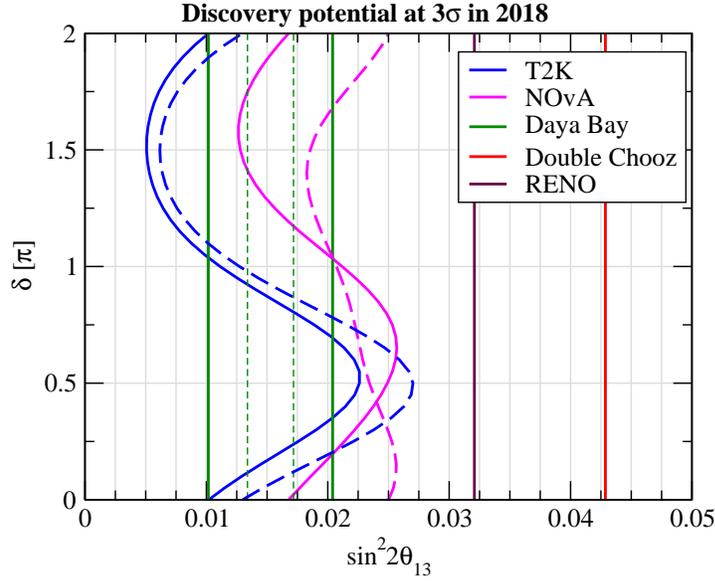}
   \caption{\label{fig:th-del} Discovery potential of the five upcoming
   experiments in the plane of $\stheta$ and $\delta$ expected in 2018, see
   section~\ref{sec:sens} for our assumptions on exposure. To the right of
   the curves a non-zero value of $\theta_{13}$ can be established at
   $3\sigma$. For the beam experiments we show normal (solid) and inverted
   (dashed) hierarchies, while reactor experiments are independent of the
   hierarchy. The four lines for Daya Bay correspond to different
   assumptions on the achieved systematic uncertainty, from weakest to
   strongest sensitivity: 0.6\% correlated among detector modules at one
   site, 0.38\% correlated, 0.38\% uncorrelated among modules, 0.18\%
   uncorrelated.}
\end{figure}

Fig.~\ref{fig:th-del} shows the $\theta_{13}$ discovery reach of the five
upcoming experiments expected in 2018. Details about our exposure
assumptions are given in section~\ref{sec:sens}. It is clear from the figure
that the discovery potential of the appearance experiments strongly depends
on the CP-phase as well as on the neutrino mass hierarchy.
We observe that the inverted hierarchy gives a weaker sensitivity. Hence, in
case no appearance signal is found the final $\theta_{13}$ limit will be set
by the IH. Note that this is completely analogous to the present information
on $\theta_{13}$ from the MINOS $\nu_e$ appearance data, compare
figs.~\ref{fig:th-del} and \ref{fig:th13-minos} (left).  The fact that the
IH has a weaker sensitivity is a consequence of the matter effect. It can be
understood by considering the $(A-1)$ terms in eq.~\ref{eq:prob-app}. Since
MINOS as well as T2K operate only with neutrinos we have $V>0$, and $A$ is
positive (negative) for NH (IH). Hence, for NH (IH) the factors $(A-1)$ in
the denominator lead to enhancement (suppression) of the oscillation
probability, and therefore IH gives a weaker limit. The different shape of
the IH curve for NO$\nu$A results from the anti-neutrino running included in
the NO$\nu$A run plan. As evident from the figure, reactor experiments are
neither sensitive to the value of $\delta$ nor to the mass hierarchy.

\subsection{On statistical errors and systematical uncertainties}
\label{sec:errors}

Apart from the differences in the oscillation physics between reactor and
super beam experiments there is also an important difference in how
statistical and systematic errors influence the final sensitivity. 

\begin{figure}
\centering \includegraphics[width=0.6\textwidth]{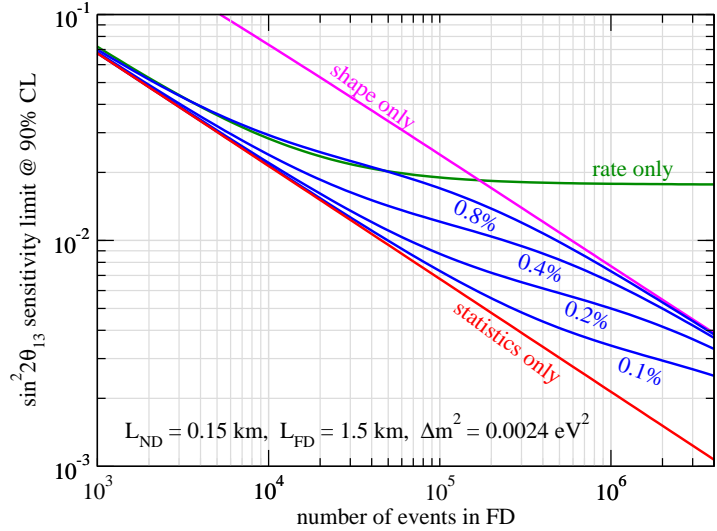}
  \caption{\label{fig:react-lumi} Illustration of the luminosity scaling of
  the $\stheta$ sensitivity at 90\%~CL of a fictitious reactor experiment.
  The horizontal axis shows the number of events in the far detector at a
  distance of 1.5~km from the reactor. We show the sensitivity for
  statistical errors only (red), as well as various values for the relative
  normalisation uncertainty between near and far detectors (blue curves with
  labels). The magenta and green curves show the sensitivity obtained for
  using only shape and only rate information, respectively.}
\end{figure}

Fig.~\ref{fig:react-lumi} shows the sensitivity to $\stheta$ as a function of
the number of events for various assumptions on systematical uncertainties for
a fictitious reactor experiment with one reactor core and two detectors, at
0.15 and 1.5~km. The statistics-only sensitivity displays the expected
$1/\sqrt{N}$ scaling. The blue curves correspond to different values of the
relative normalisation uncertainty between the two detectors. Turning on
this uncertainty deteriorates the sensitivity above a characteristic
exposure. Remarkably, at very high exposures again a $1/\sqrt{N}$ scaling is
recovered~\cite{Huber:2003pm}. This is a consequence of the information
coming from the shape of the spectrum. The magenta line in the figure shows
the limit from only shape information, with leaving the total number of
events unconstrained in the fit. In reality also the spectral shape will
suffer from some systematical uncertainty, which at some point will cut off
again the $1/\sqrt{N}$ scaling, see e.g., Refs.~\cite{Huber:2003pm,
Huber:2004ug} for a discussion. If only the total rate without any spectral
information is used, the final sensitivity is just given by the over-all
normalisation uncertainty, as illustrated by the green curve in
fig.~\ref{fig:react-lumi}.
To relate the three reactor experiments Double Chooz, RENO, and
Daya Bay with fig.~\ref{fig:react-lumi} we recall the expected
events per year given in tab.~\ref{tab:exp-summary} as $(1.5, 3, 10) \times
10^4$, respectively. Note, however, that the figure is for a fixed far
detector baseline of 1.5~km and one single reactor, which does not
correspond to any of the three experiments, and therefore the figure cannot
be applied exactly to the specific experimental configurations. 

We note that especially the high-statistics experiment Daya Bay, with let's
say $3\times 10^5$ events after 3 years, is quite sensitive to the achieved
systematic. This is illustrated in fig.~\ref{fig:th-del}, where we show the Daya
Bay sensitivity for different choices on systematics. The four lines
correspond to various assumptions on the relative normalisation uncertainty
of the 8 detector modules. 
For the most conservative limit we assume the same uncertainty as claimed by
Double Chooz, 0.6\%, and take this error correlated between detector modules
at each detector site. The Daya Bay ``baseline'' systematics is 0.38\%. The
two dashed curves correspond to this value assuming it either correlated or
uncorrelated among detector modules at one site. The most aggressive curve
shown in fig.~\ref{fig:evolsens} assumes the ``goal'' value of the
systematics of 0.18\%, uncorrelated between all detector modules. Let us
stress that of course systematics are also crucial for Double Chooz and
RENO. In those cases we have adopted a systematical error of 0.6\%, as
stated in the proposals. The final Double Chooz and RENO sensitivities will
depend in a similar way on the assumed systematics value as exemplary shown
for the case of Daya Bay in the plot.

\begin{figure}[t]
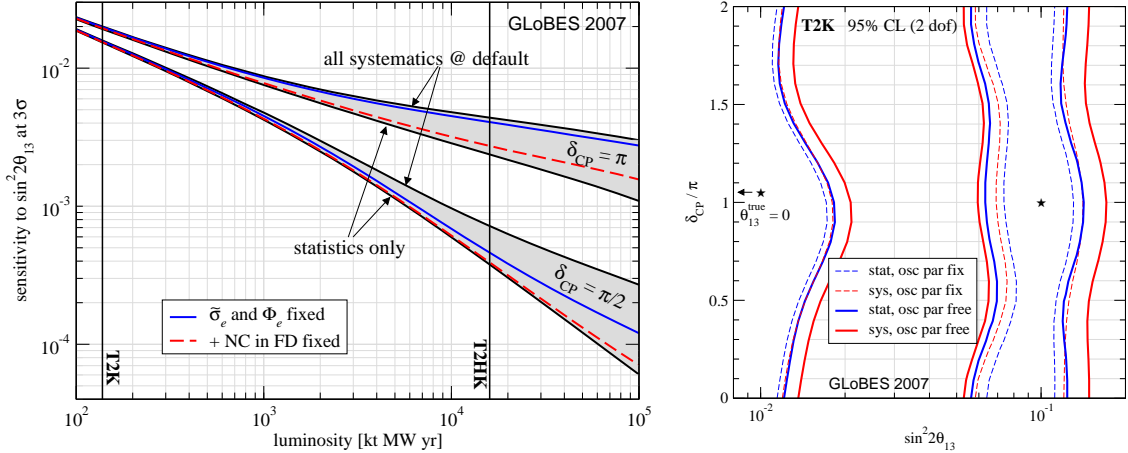
 \centering 
    \includegraphics[height=6cm]{f.17a.lumi-th13.eps}\quad
      \includegraphics[height=6cm]{f.17b.T2K-regions.eps} 
      \caption{\label{fig:lumi-th13} Left: Sensitivity to a non-zero
      $\theta_{13}$ for a T2K-like experiment as a function of
      exposure for $\delta = \pi/2$ and $\pi$ for a default choice of
      systematical errors and for statistical errors only (curves
      delimiting the shaded region).  Furthermore, the sensitivity
      obtained without uncertainty on the intrinsic beam background
      (blue) and without an uncertainty on the NC background in the
      far detector (red-dashed) is shown. Right: Allowed region in the
      plane of $\stheta$ and $\delta$ for two example choices for the
      input values marked by stars in the figure. Allowed regions are
      shown for all combinations of statistical errors only,
      systematics included, all other oscillation parameters fixed,
      and free (where for the solar parameters present errors are
      assumed). The sign($\dma$) degeneracy is neglected, and
      $\theta_{23}^\mathrm{true} = \pi/4$. Reprinted from
      Ref.~\cite{Huber:2007em}, Copyright (2008), with permission from
      JHEP.}
\end{figure}

Let us now discuss the impact of systematics and statistics on the
$\theta_{13}$ measurement in super beam experiments. As an example we
consider a sightly modified version of the T2K experiment, where, e.g., both
neutrino and anti-neutrino running is assumed with a ratio of 1:3. A
detailed description can be found in~\cite{Huber:2007em}.
Fig.~\ref{fig:lumi-th13} (left) shows the smallest value of
$\stheta$ which can be distinguished from $\theta_{13} = 0$ as a
function of the luminosity, assuming two representative values for the
CP phase which correspond roughly to the best and worst sensitivity.
The first observation is that for an exposure corresponding roughly to 
the nominal T2K exposure (marked by the vertical line labeled ``T2K'')
systematics have only a
small impact, since this measurement is largely dominated by
statistics. Numerically, the sensitivity at a T2K exposure decreases
from $\stheta = 0.0167$ to 0.0172 for $\delta = \pi/2$, and from
$\stheta = 0.0206$ to 0.0214 for $\delta = \pi$.

Only for much larger exposures systematics have a non-negligible impact on
the $\theta_{13}$ discovery reach. According to the discussion related to
eq.~\ref{eq:sig_ov_back}, the uncertainty on the background is the most
relevant systematics. Its impact is controlled by the ability of the near
detector to predict the background in the far detector.
Fig.~\ref{fig:lumi-th13} shows also curves assuming a perfectly known
$\nu_e$ beam background, and no uncertainty at all on the background ({\it
i.e.}, fixing the $\nu_e$ beam contamination as well as the NC background).
If the total background is fixed the sensitivity is close to the pure
statistics case. It is interesting to note that for the two examples of
$\delta$ shown in the figure the importance of beam and NC backgrounds is
different. This is an effect of the spectral shapes of the signal relative
to the background, since the spectrum of the signal depends on the value of
$\delta$, and also beam and NC backgrounds have rather different shapes.

Fig.~\ref{fig:lumi-th13} (right) shows the allowed region in the space of
$\stheta$ and $\delta$ obtained by the same modified T2K-like
configuration.\footnote{Note the different shapes of the regions from
fig.~\ref{fig:lumi-th13} (right) and the corresponding panels of
fig.~\ref{fig:deltatheta}. The reason is the anti-neutrino running assumed
for fig.~\ref{fig:lumi-th13}.} The impact of systematics is small, though
not negligible in this case. Furthermore, we observe that the uncertainty on
the other oscillation parameters has a sizable impact on the allowed region.
This effect comes entirely from the atmospheric parameters $\dma$ and
$\theta_{23}$. Apparently the disappearance channel does not provide enough
accuracy on these parameters to avoid an effect on the $\theta_{13}$
determination. For the solar parameters the accuracy from present data is
sufficient to eliminate any effect on the results shown in the figure.

\subsection{Sensitivity predictions for the upcoming experiments}
\label{sec:sens}

We now discuss the sensitivity of the different experiments to
$\theta_{13}$, using two different performance indicators: the $\theta_{13}$
sensitivity limit and the $\theta_{13}$ discovery potential. The
$\theta_{13}$ sensitivity limit describes the ability of an experiment to
constrain $\theta_{13}$ if no signal is seen. It is basically determined by
the worst case parameter combination which may fake the simulated
$\theta_{13}=0$. 
The sensitivity limit does not depend on the simulated hierarchy and
$\delta$, as $\theta_{13}=0$ is used to calculate the fake data. For a more
detailed discussion, see appendix~C of~\cite{Huber:2004ug}. The $\theta_{13}$
discovery potential is given by the smallest true value of $\theta_{13} > 0$
which cannot be fitted with $\theta_{13}=0$ at a given CL. Since the
simulated $\theta_{13}$, $\delta$, and hierarchy determine the simulated
rates, the $\theta_{13}$ discovery potential will depend on the values of
all these parameters chosen by nature. On the other hand, correlations and
degeneracies are of minor importance because for the fit $\theta_{13}=0$ is
used.

For reactor experiments both measures (sensitivity limit and discovery
potential) are very similar, since statistics are Gaussian and the
oscillation physics is simple. For beam experiments the smallest
$\theta_{13}$ discovery potential for all values of $\delta$ and mass
hierarchies (risk-minimised $\theta_{13}$ discovery potential) is often
similar to the $\theta_{13}$ sensitivity limit. However, notable deviations
from this rule occur due to Poisson statistics as well as more complicated
oscillation physics implying correlations and degeneracies. In particular,
we find that for low exposures the worst case discovery potential of the
beams is somewhat better than the limit in case of no signal at the same
confidence level.

We are going to show the evolution of the $\theta_{13}$ sensitivity
with time, assuming the following schedules for the experiments based
on up-to-date information. For Double Chooz the far detector starts
data taking in June 2010, while near detector data will become
available beginning of 2012. RENO starts data taking with both close
and far detectors end of 2010.  For Daya Bay we do not consider the
periods where only the near detectors are available in 2011, but
consider the start of the full experiment with all detectors at the
end of 2011. For all reactor experiments we assume that all reactors
are at nominal power all time according to the event numbers per year
given in table~\ref{tab:exp-summary}. For T2K and NO$\nu$A we use the
exposures as a function of time shown in fig.~\ref{fig:beam-exposure},
assuming that protons on target are uniformly distributed along the
year, not taking into consideration the specific schedules of the
accelerators.  In all cases we assume that results are available
instantaneously with data.

The simulations of T2K and NO$\nu$A are performed with the GLoBES
software~\cite{Huber:2004ka, Huber:2007ji} based on the experiment
definitions developed in~\cite{Huber:2009cw} and available at the
GLoBES web-page. Modifications to the T2K simulation have been
introduced due to recent updates on efficiencies (it now reproduces
fig.~\ref{fig:T2K-events}).  For the reactor experiments an
independent code has been developed, allowing for an arbitrary number
of reactors and detectors. Various systematics are included with
proper correlations between detectors and reactors, as well as
backgrounds from accidental, fast neutrons, and cosmogentics according
to the numbers provided in the respective proposals. Reactor fluxes
and their uncertainties are included following~\cite{Huber:2004xh}.

\begin{figure}[t]
\begin{center}
\includegraphics[width=0.6\textwidth]{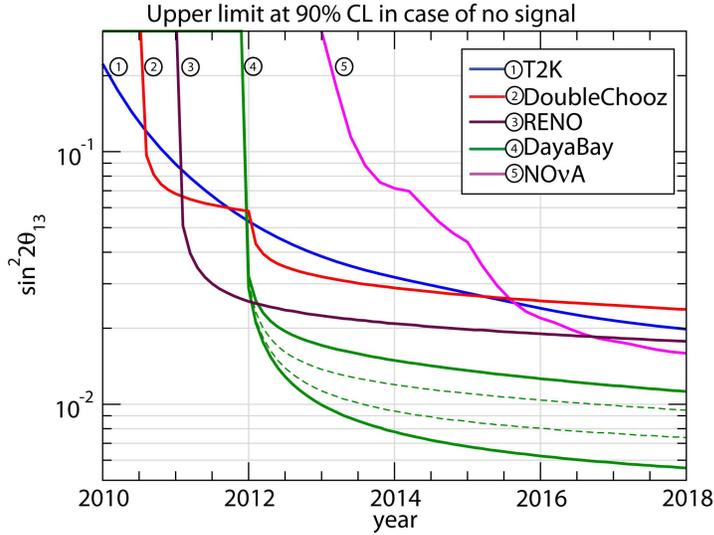}
\end{center}
\caption{\label{fig:evolsens} Evolution of the $\theta_{13}$ sensitivity
  limit as a function of time (90\% CL), i.e., the 90\%~CL limit which will
  be obtained if the true $\theta_{13}$ is zero. The four curves for Daya
  Bay correspond to different assumptions on the achieved systematic
  uncertainty, from weakest to strongest sensitivity: 0.6\% correlated among
  detector modules at one site, 0.38\% correlated, 0.38\% uncorrelated among
  modules, 0.18\% uncorrelated.}
\end{figure}

The $\theta_{13}$ sensitivity limit time evolution is shown in 
fig.~\ref{fig:evolsens}. We observe that the global sensitivity limit will
be dominated by reactor experiments. If the assumed schedules for the
reactor experiments are achieved, Double Chooz and RENO will each dominate
the limit between 0.5 and one year. As soon as Daya Bay becomes operational
it will have the best limit (as well as discovery potential) among the
reactor experiments, thanks to the large exposure. After 5 years of running
the limits at 90\%~CL will be $\stheta = 0.026$ and 0.019 for Double Chooz
and RENO, respectively, whereas the Daya Bay limit will range from 0.012 to
0.006, depending on the systematics. In fig.~\ref{fig:evolsens} we show the
Daya Bay limit under various assumptions about the relative normalisation
uncertainty of their detectors, as discussed in section~\ref{sec:errors}.
Even for the most pessimistic assumption, Daya Bay will set the final limit
on $\theta_{13}$ in the case no finite value will be discovered. 

\begin{figure}
  \centering
  \includegraphics[width=0.6\textwidth]{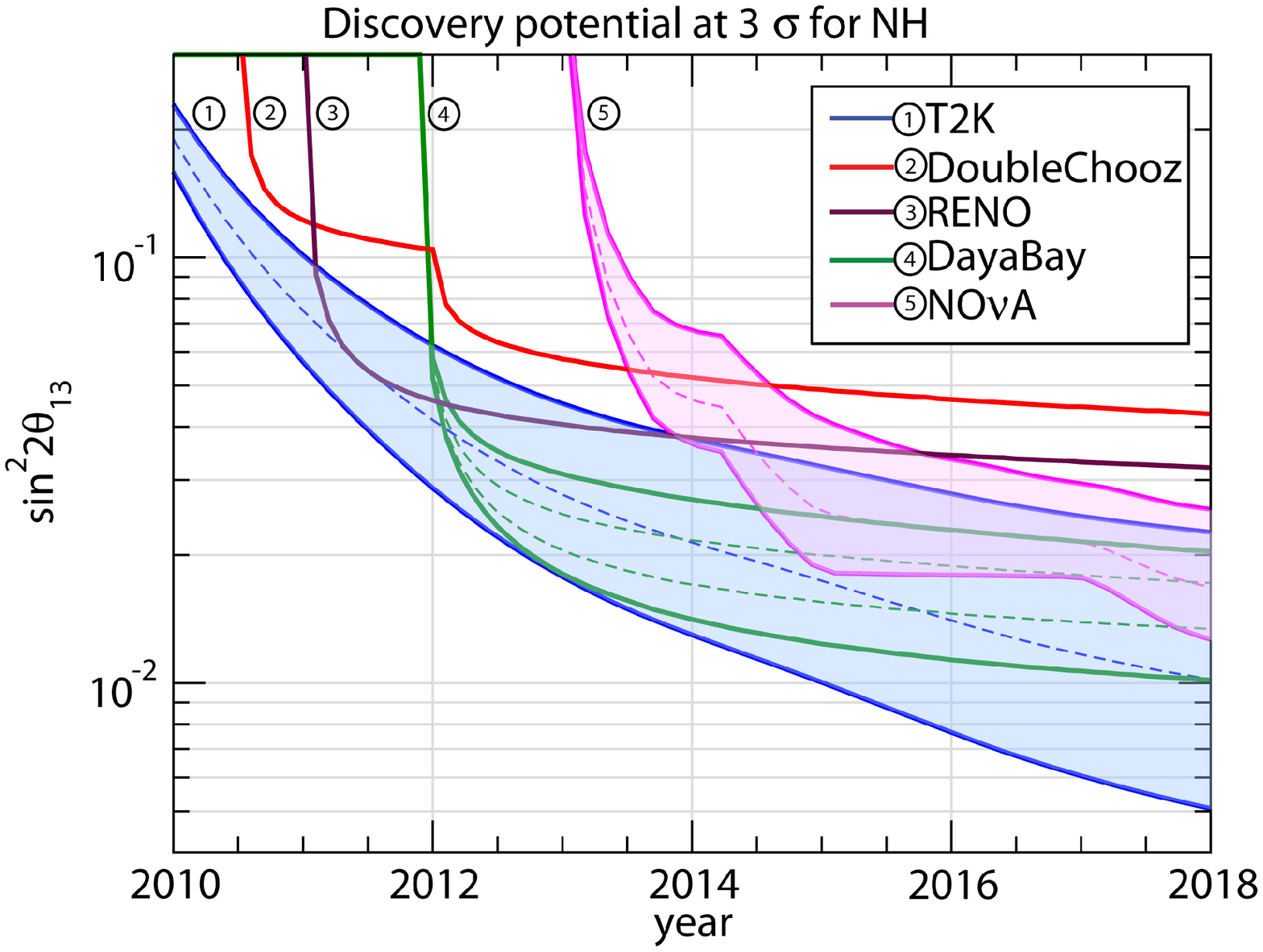}\\
  \includegraphics[width=0.6\textwidth]{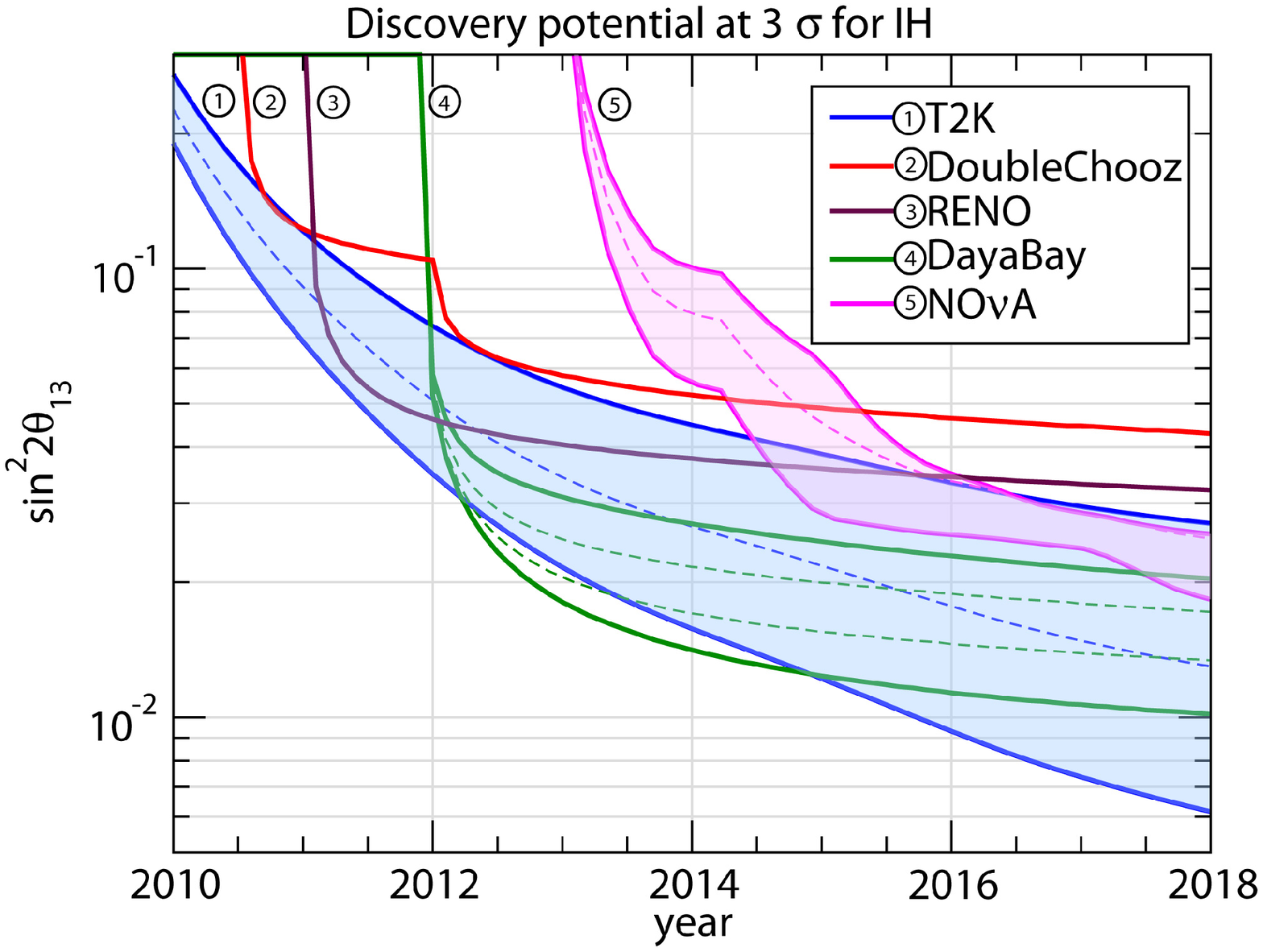}
  \caption{\label{fig:evoldisc} Evolution of the $\theta_{13}$ discovery
  potential as a function of time ($3\sigma$~CL), i.e., the smallest value
  of $\theta_{13}$ which can be distinguished from zero at $3\sigma$. We
  assume the normal and inverted simulated hierarchies in the top and bottom
  panels, respectively.  The bands for the beams reflect the (unknown) true
  value of $\delta$. For the dashed curves $\delta = 0$ has been fixed.  The
  four curves for Daya Bay correspond to different assumptions on the
  achieved systematic uncertainty, from weakest to strongest sensitivity:
  0.6\% correlated among detector modules at one site, 0.38\% correlated,
  0.38\% uncorrelated among modules, 0.18\% uncorrelated.}
\end{figure}

In case of no signal, the $\theta_{13}$ limit from beam experiments suffers
from the marginalization over the CP phase and the mass hierarchy, as
discussed in section~\ref{sec:complementary}. This situation is very different
in case of the discovery potential, since there a favourable value of
$\delta$ can greatly enhance the sensitivity of the appearance experiments. 
The $\theta_{13}$ discovery potentials are shown in fig.~\ref{fig:evoldisc}
as a function of time. For the beam experiments, the dependence on the true
value of $\delta$ is reflected by the interval between the solid curves for
a given time (shaded regions). The dashed curves for T2K and NO$\nu$A
correspond to a fixed value for the CP phase of
$\delta=0$.~\footnote{Evolution of sensitivities under this condition have
been shown recently in \cite{Mezzetto:2009cr, Suzuki:2010zz}.} The reactor
experiments are not affected by the true $\delta$; the various curves for
Daya Bay again correspond to the different assumptions concerning
systematics as described above. There is a small dependence on the true mass
hierarchy for the beam experiments, compare top and bottom panels, where for
IH the sensitivity is slightly worse because of the suppression of the
oscillation probability for neutrinos by the matter effect.

The comparison of Figs.~\ref{fig:evoldisc} and \ref{fig:evolsens} shows that
suitable values of $\delta$ may significantly improve the discovery
potential of beams compared to their sensitivity limit. Indeed, T2K may
discover $\theta_{13}$ for smaller $\theta_{13}$ than Daya Bay in a
significant fraction of the parameter space, depending on the achieved
systematics in Daya Bay. The NO$\nu$A band becomes more narrow due to the
complementary information from the anti-neutrino running, with the clear
disadvantage of being somewhat late.

\begin{figure}
  \centering
  \includegraphics[width=0.6\textwidth]{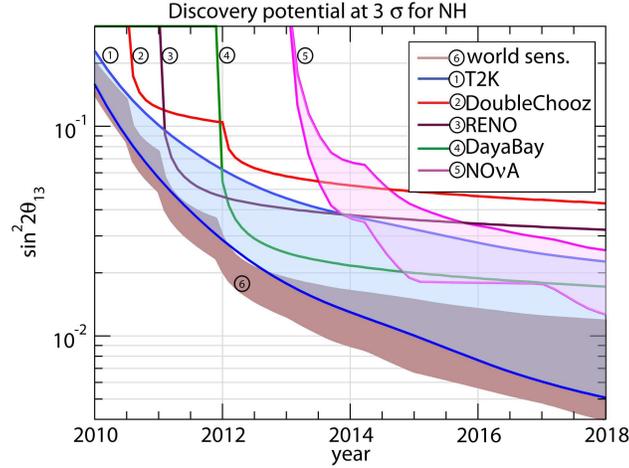}
  \caption{\label{fig:evoldisc-global} Evolution of the $\theta_{13}$
  discovery potential as a function of time ($3\sigma$~CL) for NH, showing
  the global sensitivity reach. The bands for the beams and the global reach
  reflect the (unknown) true value of $\delta$. For Daya Bay we have assumed
  a systematical uncertainty of 0.38\% correlated among detector modules at
  one site.}
\end{figure}

In figure~\ref{fig:evoldisc-global} we illustrate how the world sensitivity
to $\theta_{13}$ could look like under the assumptions of the above
schedules and that at each point in time a combined analysis of all
available data is performed. The discovery reach will be set roughly by the
optimal sensitivity of T2K, where the reactor experiments play an important
role in providing sensitivity for the values of $\delta$ unfavourable for
T2K. This plot nicely illustrates the interplay between reactor and beam
experiments and shows that the global reach can be enhanced significantly if
experiments of both types are available simultaneously with comparable
sensitivities.

Let us comment also on the dependence of the sensitivities on the assumption
$\sin^2\theta_{23} = 0.5$. This is relevant only for beam experiments, since
the survival probability in the reactor experiments is independent of
$\theta_{23}$, see eq.~\ref{eq:prob-react}. In the range $0.4 <
\sin^2\theta_{23} < 0.6$ (corresponding roughly to the current $2\sigma$
allowed range~\cite{Schwetz:2008er}) the effect for the beam experiments is
small~\cite{Huber:2009cw}. The sensitivity to $\theta_{13}$ becomes somewhat
better the larger $\sin^2\theta_{23}$, which follows from the first term in
eq.~\ref{eq:prob-app}. In these cases the choice $\sin^2\theta_{23}=0.5$
corresponds to the ``average'' situation. 

Note that this discussion is based on the unitary standard three--flavour
oscillation framework. If the search for new physics is taken into account,
different reactor experiments, or reactor experiments and superbeams, may
imply different information and therefore be very complementary; see, e.g.,
Refs.~\cite{Schwetz:2005fy, Kopp:2007ne}.

\subsection{On the interpretation of sensitivities for future experiments}

A widely used procedure to calculate sensitivities for future experiments is
to assume some input values (``true'' or ``simulated'' values) for the
oscillation parameters for which the predictions for the observables in a
given experiment are calculated without statistical fluctuations. Then these
predictions are used as ``data'' and a statistical analysis of these data is
performed to see how well the input values for the parameters can be
reconstructed by the experiment. This procedure, denoted by ``standard''
procedure in the following, should give the sensitivity of an ``average''
experiment, where ``average'' lacks a precise definition.
The results of the previous subsections are obtained by this method, and
most of the sensitivity studies for future neutrino oscillation experiments in
the literature are based on this procedure. In particular the GLoBES
long-baseline experiment simulation software~\cite{Huber:2004ka,
Huber:2007ji} is designed for this method.

Ref.~\cite{Schwetz:2006md} clarifies the correct interpretation of such
sensitivities in the context of oscillation experiments. Monte Carlo
simulations of the Double Chooz and T2K experiments are performed in order
to address the following question in a frequentist framework: {\it Given a
true value of $\theta_{13}$, what is the probability that the hypothesis
$\theta_{13} = 0$ can be excluded at a certain confidence level?} 
This generalises the usual sensitivity limits to a well defined
statistical statement and allows also a precise definition of the
``average experiment''. For example one may define the
sensitivity of an average experiment as the value of
$\theta_{13}^\mathrm{true}$ for which $\theta_{13}=0$ can be excluded
with a probability of 50\%. 

A large number of artificial data sets is generated to calculate the actual
distribution of the statistics used to decide whether $\theta_{13}=0$ should
be rejected at a given confidence level. This allows to answer the question
stated above within a well defined frequentist framework.
Moreover one does not rely on questionable assumptions necessary in
the standard procedure, for example issues related to the non-linear
character of the parameters, the periodicity of the CP phase
$\delta$, the physical boundary $\stheta \ge 0$, assuming standard
$\chi^2$-distributions, and the question of how many degrees of
freedom to use for them.
Note that the specific experimental configurations and parameters used for
the simulations in~\cite{Schwetz:2006md} for Double Chooz and T2K differ
slightly from the most up-to-date versions.

The probability $P_\mathrm{disc}(\alpha, \theta_{13})$ that $\theta_{13}=0$
can be excluded at the $100(1-\alpha)\%$~CL is given by
\begin{equation}\label{eq:PdiscFC}
P_\mathrm{disc} (\alpha, \theta_{13})\equiv 
P\left[\Delta\chi^2_0 > \lambda(\alpha) \: | \: \theta_{13}\right]
= \int_{\lambda(\alpha)}^\infty dx\, f_{\theta_{13}}(x) \,.
\end{equation}
Here $\Delta\chi^2_0 \equiv \chi^2(\theta_{13} = 0) - \chi^2_\mathrm{min}$
is the difference in $\chi^2$ between the best fit point and $\theta_{13} =
0$, $\lambda(\alpha)$ is the $\Delta\chi^2$ value corresponding to the
$100(1-\alpha)\%$ confidence level, and $f_{\theta_{13}}(\Delta \chi^2_0)$
is the distribution of $\Delta\chi^2_0$ for a fixed true value of
$\theta_{13}$. The distribution $f_{\theta_{13}}(\Delta \chi^2_0)$ and the values of
$\lambda(\alpha)$ have been obtained by Monte Carlo simulation.

\begin{figure}
\centering 
\includegraphics[width=0.6\textwidth]{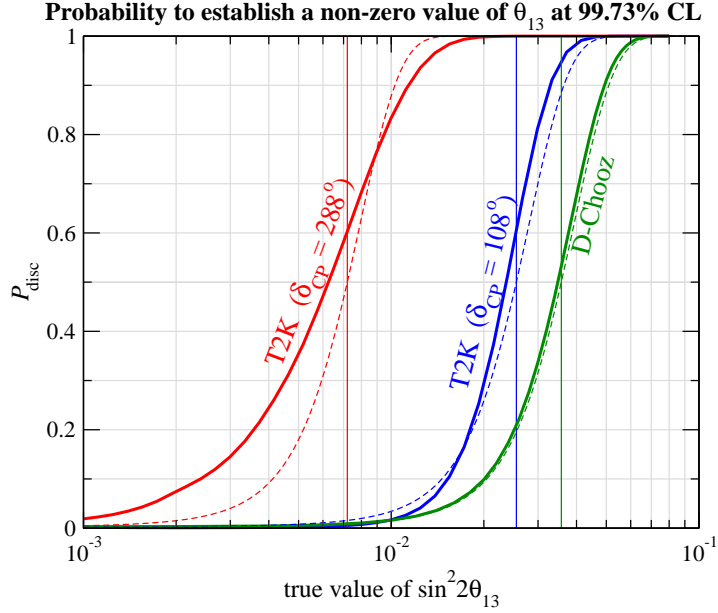}
  \caption{The probability to exclude the hypothesis $\theta_{13} = 0$
  at the 99.73\%~CL for T2K and Double Chooz as a function of the true
  value of $\stheta$. The two curves for T2K correspond to the true
  values $\delta = 108^\circ$ and $288^\circ$. The vertical lines show
  the corresponding ``standard'' sensitivities. The dashed curves
  correspond to the probability $P_\mathrm{disc}$ calculated in the
  Gaussian approximation according to eq.~\ref{eq:gauss}. Reprinted
  from Ref.~\cite{Schwetz:2006md}, Copyright (2007), with permission
  from Elsevier.}
\label{fig:th13-prob}
\end{figure}

Fig.~\ref{fig:th13-prob} shows the probability
$P_\mathrm{disc}$ to exclude the hypothesis $\theta_{13} = 0$ at the
99.73\%~CL for T2K and Double Chooz as a function of the true value of
$\stheta$. For each true value $3\times 10^6$ data sets have been
simulated. For T2K the two values chosen for $\delta$ correspond
roughly to the best and worst sensitivity. The vertical lines in the
plot show the standard sensitivities calculated from the condition
$\Delta \chi^2 \ge 9$ without statistical fluctuations. One observes
that for Double Chooz the standard sensitivity corresponds indeed with good
accuracy to $P_\mathrm{disc} = 50\%$, as expected for an ``average''
experiment. For T2K the discovery probabilities corresponding to the
standard sensitivities are actually slightly higher, around 60\%.

The dashed curves shown in fig.~\ref{fig:th13-prob} are obtained
assuming a Gaussian measurement of $\stheta$. In this case
$P_\mathrm{disc}$ can be obtained in terms of the error function in the
following way. Assuming that $x$ is a Gaussian variable with standard
deviation $\sigma$ the hypothesis $x = 0$ can be excluded at the
99.73\%~CL if the observed value $x^\mathrm{obs}$ is bigger than
$3\sigma$. On the other hand, the probability for $x^\mathrm{obs} \ge
3\sigma$ as a function of the true value $x^\mathrm{true}$ is easily
calculated as
\begin{equation}\fl\label{eq:gauss}
P\left[ x^\mathrm{obs} \ge 3\sigma \:|\: x^\mathrm{true} \right] =
\int_{3\sigma}^\infty dx \, G(x; \: x^\mathrm{true}, \sigma) =
\frac{1}{2} \left[ 1 - \mathrm{erf}
\left(\frac{3\sigma - x^\mathrm{true}}{\sqrt{2}\sigma}\right)\right] \,,
\end{equation}
where $G(x; \: x^\mathrm{true}, \sigma)$ denotes the normal
distribution with mean $x^\mathrm{true}$ and standard deviation
$\sigma$. 

The dashed curves in fig.~\ref{fig:th13-prob} have been obtained from
eq.~\ref{eq:gauss} by identifying $\stheta = x$ and by using for
$\sigma$ one third of the 99.73\%~CL sensitivity limit from the
standard procedure. One observes that for Double Chooz this approximation
is excellent. Hence, in this case $\stheta$ can be considered indeed
as a Gaussian variable and the probability $P_\mathrm{disc}$ can be
calculated from the standard sensitivity limit and
eq.~\ref{eq:gauss} without the need of a MC simulation.
In contrast, for T2K some deviations from Gaussianity are visible
(especially for $\delta^\mathrm{true} = 288^\circ$). This is not
unexpected, since in this case event numbers are small, background
fluctuations are important, and the dependence of the observables on
the parameters is much more complicated than in the case of Double Chooz.

\begin{figure}
\centering 
\includegraphics[width=0.65\textwidth]{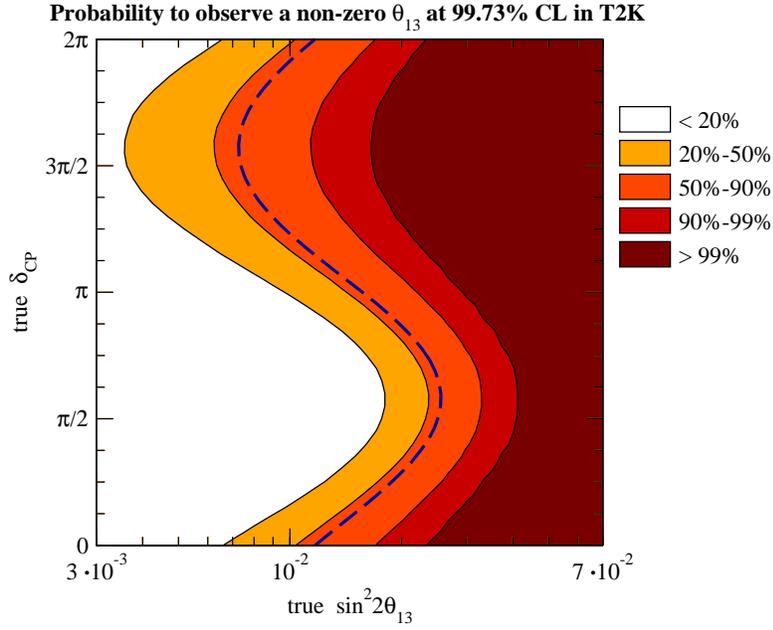}
  \caption{Contours of the probability $P_\mathrm{disc}$ to establish
  a non-zero value of $\theta_{31}$ at the 99.73\%~CL for T2K in the
  $\stheta^\mathrm{true}$-$\delta^\mathrm{true}$ plane.  The dashed
  curve corresponds to the ``standard sensitivity limit''.  Reprinted
  from Ref.~\cite{Schwetz:2006md}, Copyright (2007), with permission
  from Elsevier.}
\label{fig:th13-contours}
\end{figure}

Contours of the probability $P_\mathrm{disc}$ for the T2K experiment
in the plane of $\stheta^\mathrm{true}$ and $\delta^\mathrm{true}$
are shown in fig.~\ref{fig:th13-contours}. $P_\mathrm{disc}$ has been
calculated for a grid of $41\times 41$ values and at each point in the
grid $10^5$ data sets have been generated, leading in total to nearly
$1.7\times 10^8$ performed fits. This figure is the generalisation of
the usual sensitivity limit (shown as dashed curve) and for each true
value of the parameters one can infer the probability that T2K can
establish a non-zero value of $\theta_{13}$ at the 99.73\%~CL. As
indicated already in fig.~\ref{fig:th13-prob} one finds that the
standard sensitivity curve corresponds roughly to a discovery
probability of 60\%. The region where $\theta_{13} > 0$ can be
established with high probability, let's say greater than 99\%, is
found for $\stheta^\mathrm{true} > 0.0166 - 0.041$, depending on the
true value of $\delta$. It is shifted with respect to the standard
sensitivity limit to values of $\stheta$ larger by roughly a factor of
2.

In summary, for Double Chooz (and reactor experiments in general) the
Gaussian approximation is very well justified. The usually calculated
sensitivity corresponds to the performance of an average experiment
(the discovery will be made with a probability of 50\%), and the
actual discovery probability can be estimated by a simple formula in
terms of the error function. In the case of appearance beam
experiments some deviations from Gaussianity are found. For T2K the
standard sensitivity limits correspond to a discovery probability of
about 60\%.
The results of~\cite{Schwetz:2006md} confirm that standard sensitivity
limits provide a reasonable approximation for an average experiment,
corresponding to a discovery probability of order 50\%. However, one has to
be aware of the correct interpretation of such limits. In general the region
where a discovery can be made with high probability is significantly smaller
than the one corresponding to the standard sensitivity limits.

\section{A subsequent generation of experiments\label{sec:next}}
\label{sec:future}

If the upcoming generation of experiments discussed in this review cannot
establish a finite value of $\theta_{13}$, experiments able to explore the
region $\stheta < 10^{-2}$ will be needed. On the other hand, if a positive
signal can be established this generation of experiments will not be
sensitive enough to make a firm discovery (3$\sigma$ or better) of leptonic
CP violation and the neutrino mass hierarchy, see Fig.~\ref{fig:deltatheta}
and the extensive discussion in Ref.~\cite{Huber:2009cw}. In this section we
briefely outline some possibilities for such high precission oscillation
facilities, more complete discussions can be found in
\cite{Bandyopadhyay:2007kx, Bernabeu:2010rz}.


A possible way to attack the searches of leptonic CP violation is to
push conventional neutrino beams to their ultimate performances and
build huge detectors, one order of magnitude, at least, bigger than
the present detectors.
As an example T2K could be upgraded, as already delineated in the initial
LoI~\cite{Itow:2001ee}, by pushing the power of the J-PARC Main Ring to 4~MW
(2~MW seems nowadays more realistic) and building a water \ceren detector of
about 500~kton fiducial volume, HyperKamiokande, more than 20 times bigger
than SuperKamiokande, to be placed at the same distance and the same
off-axis angle as SuperKamiokande. Subsequent developments of this project
foresee the possibility of placing half of the detector at a longer
distance, about 900~km in Korea, T2KK~\cite{T2KK}, or to substitute it with
a 100~kton liquid argon detector placed at 658~km from J-PARC at a smaller
off-axis angle of $0.5^\circ$~\cite{Badertscher:2008bp}. The longer
baselines would provide better sensitivities to the mass hierarchy and
better control of degeneracies.
In the United States a similar project designs a wide-band beam
(WBB)~\cite{Diwan:2006qf} generated by the FNAL Main Injector and fired to a
a 300~kton water \ceren detector placed at the DUSEL lab, 1290~km far away
from FNAL ($3 \div 6$ liquid argon modules of 20~kton are also taken in
consideration). 

In Europe, superbeams have been proposed based on upgrades of the CNGS beam
\cite{Meregaglia:2006du, Baibussinov:2007ea}, or on a high power version of
accelerators foreseen for a possible new injection chain of the
LHC~\cite{LHC-Upgrade} as the SPL and the PS2.
It appears very difficult~\cite{meddahi} to set the performances of the CNGS
beam to the level needed for searches of leptonic CP violation, that is 10 times
the protons actually generated on target.
The SPL superbeam \cite{GomezCadenas:2001eu}, based on the 3.5~GeV, 4~MW
Superconducting Proton Linac~\cite{SPL}, can generate a neutrino beam of
about 300~MeV sent to a 500~kton water \ceren detector (MEMPHYS
\cite{Memphys}) placed in Fr\'ejus at 130~km from CERN. It would have
excellent sensitivities for \thetaot and leptonic CP
violation~\cite{Campagne:2006yx}, while the mass hierarchy sensitivity would
be limited by the relatively short baseline (the synergy with atmospheric
neutrinos~\cite{Huber:2005ep} would provide sensitivity to the mass
hierarchy at large values of \thetaot \cite{Campagne:2006yx}).

It has been proposed in~\cite{Rubbia:2010fm} to generate a neutrino beam by
a high power (1.6~MW) version of the PS2 accelerator, a 50~GeV synchrotron
designed to run at 0.4~MW to serve as a component of the new injection
scheme for the LHC. Neutrinos could be then fired to a 100~kton liquid argon
detector, placed at distances between 950~km or 1544~km or 2300~km. The
distances correspond to the three underground labs of Sieroszowice in
Poland, Slanic in Romania and Pyhasalmi in Finland, respectively, 
considered by the LAGUNA FP7 Design Study \cite{LagunaDS}. As in the case
of the WBB at DUSEL, this setup would measure neutrinos at the first and at
the second oscillation maximum. Liquid argon is certainly the best candidate
to fulfill the requirements of this configuration. This kind of
configuration would have excellent performances in measuring the sign of
$\dma$ but a limited sensitivity for leptonic CP violation and the
measurement of \thetaot.

\begin{figure}
  \centering \includegraphics[width=0.8\textwidth]{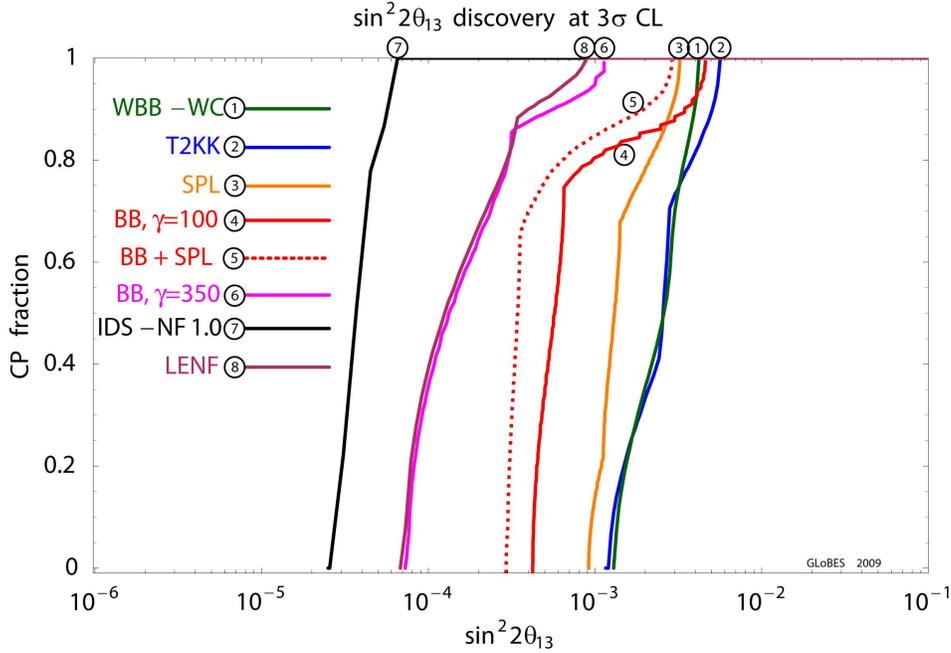}
   \caption{$3\sigma$ discovery sensitivity to $\stheta$ as a function of
   the fraction of all possible values of \delCP, computed for 10 years of
   data taking and 5\% systematic errors (2\% for the neutrino factory).
   The curve labelled WBB-WC refers to the FNAL--DUSEL setup
   \protect\cite{Barger:2007jq}, the curves for the SPL superbeam, the
   $\gamma=100$ beta beam and their combination are taken from
   \protect\cite{Campagne:2006yx}, the curve for the $\gamma=350$ beta beam
   is taken from \protect\cite{Choubey:2009ks}, the low energy neutrino
   factory (LENF) from \protect\cite{Bross:2007ts}, and IDS-NF corresponds
   to a hypothetical neutrino factory scenario with two 100~kton magnetized
   iron detectors at 4000 and 7500~km and an Emulsion Cloud Chamber detector
   to detect $\nue \to \nu_\tau$ transitions at 4000~km taken from
   \protect\cite{Bernabeu:2010rz}.  The plot has been adapted from a figure
   prepared by P.~Huber for the EURO$\nu$ WP6 study
   \protect\cite{Bernabeu:2010rz}. \label{fig:th13-future}}
\end{figure}

Sensitivities on \thetaot of these future superbeams could reach the
$\sin^2(2\thetaot) \simeq 10^{-3}$ level, as illustrated in
Fig.~\ref{fig:th13-future}. Similarly also leptonic CP violation could be
performed in the same range of \thetaot values, see
e.g.,~\cite{Bandyopadhyay:2007kx, Barger:2007yw, Bernabeu:2010rz}.

The main limitations of conventional neutrino beams are the intrinsic
\nue contamination of the beam produced by the decay-in-flight of the
kaons produced at the target and by the muons from the pion decays,
and the uncertainty of the beam flux prediction
due to the finite precision with which the hadron production
cross sections can be known.
The intrinsic limitations of conventional neutrino beams can be overcome if
the neutrino parents are fully selected, collimated and accelerated to a
given energy. This can be attempted within the muon lifetime (neutrino
factory~\cite{Geer:1997iz}) or within beta decaying ion lifetimes (beta
beam~\cite{Zucchelli:2002sa}). In this way very pure, intense and precisely
known neutrino beams can be designed.

The beta beam~\cite{Zucchelli:2002sa} is a facility based on the decay in
flight of $\beta$-unstable ions, for reviews see \cite{Volpe:2006in,
Lindroos:2010zza}. They are ideal tools to study $\nue\to\numu$ transitions
and their CP-conjugate without any intrinsic source of backgrounds.
The original proposal of Ref.~\cite{Zucchelli:2002sa} was tuned to leverage
at most the present facilities of CERN (the PS and the SPS) and it was based
on $^6$He and $^{18}$Ne as \nubare and \nue sources, respectively. These ions
are accelerated to $\gamma=100$~\cite{Mezzetto:2005ae} and stored in a decay
ring of about 7~km. With MEMPHYS as a far detector at Fr\'ejus, this set-up
could outperform any superbeam configuration in terms of sensitivity to
\thetaot and leptonic CP violation \cite{Campagne:2006yx}, and these
performances would be significantly enhanced if the SPL superbeam would be
fired to the same far detector \cite{Mezzetto:2005yf, Campagne:2006yx}.
Accelerating the same ions to $\gamma=350$ (an option that requires a 1~TeV
accelerator) the performances of the beta beam would be significantly
improved \cite{BurguetCastell:2003vv, BurguetCastell:2005pa}. Various even
more ambitious beta beam configurations are discussed in the literature,
including very high $\gamma$ factors as well as exploring different
isotopes, see~\cite{Bernabeu:2010rz} for a recent summary and references.

 
Production, acceleration and stacking of high intensity muon beams for
muon colliders have been envisaged thanks to 
their decays producing useful beams of
$\numu$ and $\nubare$ (exploiting $\mu^-$ decays into $e^- \nubare
\numu$) or $\nubarmu$ and $\nue$ ($\mu^+$ decays into $e^+ \nue
\nubarmu$). 
In a ``Neutrino Factory''~\cite{Geer:1997iz}, muons are created from an
intense pion source at low energies, their phase space is compressed to produce
a bright muon beam, which is then accelerated to the desired energy and injected
into a storage ring with long straight sections pointing in the desired
direction. The neutrino factory design can be considered as strongly
physics-motivated intermediate step towards a muon collider
\cite{Geer:2009zz}. 

Assuming $\mu^+$ in the decay ring, one looks for $\nue\to\numu$
oscillations due to the appearance of $\mu^-$ from $\numu$ CC events in the
detector (``wrong sign muons''), which have to be separated from the bulk of
$\mu^+$ (``right sign muons'') coming from unoscillated $\nubarmu$. A
suitable detector to search for these transitions is a magnetized iron
detector~\cite{Mind, Cervera:2010rz}.
As pointed out in~\cite{DeRujula:1998hd}, a neutrino factory would be an
ideal tool to address CP violation in the leptonic sector, with outstanding
performances compared with pion-based sources. The realization of the
neutrino factory still represents a major accelerator challenge compared
with neutrino superbeams.

We report in Fig.~\ref{fig:th13-future} a comparison of \thetaot
sensitivities of some examples for the facilities described in
this section.

\section{Conclusions}
\label{sec:conclusions}

In this review we have summarized the present status of the last unknown
lepton mixing angle $\theta_{13}$ and discussed the experimental progress to
be expected within the next years. Although there some hints for a non-zero
value of $\theta_{13}$ in current neutrino oscillation data, at present we
can only quote an upper bound of $\sin^2\theta_{13} < 0.031$ at 90\%~CL from
global data. Best fit points obtained by different groups range from
$\sin^2\theta_{13} \simeq 0.01$ to $0.02$, being consistent with
$\theta_{13}=0$ at the 1 to 2$\sigma$ level, see
Tab.~\ref{tab:fit-comparison} for a summary.

Several new neutrino oscillation experiments will contribute significantly
to our information on $\theta_{13}$ in the near future. The reactor
experiments Daya Bay, Double Chooz, and RENO as well as the accelerator
experiments NO$\nu$A and T2K have been presented in section~\ref{sec:exp}.
The time evolution for the sensitivity to $\theta_{13}$ from these
experiments is summarised in the figures~\ref{fig:evolsens} and
\ref{fig:evoldisc}. Within the next few years values of $\stheta$ down to
$10^{-2}$ will be probed, about one order of magnitude smaller than the
current upper bound. These results will be of fundamental importance for the
better understanding of the lepton sector, the problem of flavour and in
particular for the question whether the values of the lepton mixing matrix
indicate the presence of a flavour symmetry. The results of those
experiments will be a corner stone for any subsequent neutrino oscillation
facility aiming at the ultimate goals like the discovery of CP violation in
the lepton sector or the determination of the neutrino mass hierarchy.

\section*{Acknowledgements}

We thank Mark Messier for information on the NO$\nu$A experiment, 
Soo-Bong Kim for information on the RENO experiment and
Kam-Biu Luk for information on the Daya Bay experiment.
We acknowledge the financial support of the European Community under the
European Commission Framework Programme 7 Design Studies: EUROnu, Project
Number 212372 and LAGUNA, Project Number 212343. The EC is not liable for
any use that may be made of the information contained herein.
The work of TS is partly supported by the Transregio
Sonderforschungsbereich TR27 ``Neutrinos and Beyond'' der Deutschen
Forschungsgemeinschaft.

\newpage

\section*{References}

\bibliographystyle{my-h-physrev}
\bibliography{./th13}

\end{document}